\documentstyle[12pt,epsf]{article}
\begin{document}

\title{Gauge dependence of effective gravitational field}

\author{~Kirill~A.~Kazakov\thanks{Email address: kirill@theor.phys.msu.su} ~and
~Petr~I.~Pronin\thanks{Email address: petr@theor.phys.msu.su}}

\maketitle

\begin{center}
{\em Moscow State University, Physics Faculty,\\
Department of Theoretical Physics.\\
$117234$, Moscow, Russian Federation}
\end{center}

\begin{abstract}
The problem of gauge independent definition of effective gravitational
field is considered from the point of view of the process of measurement.
Under assumption that dynamics of the measuring apparatus can be described
by the ordinary classical action, effective Slavnov identities for the
generating functionals of Green functions corresponding to a system of
arbitrary gravitational field measured by means of scalar particles
are obtained. With the help of these identities, the total gauge dependence
of the non-local part of the one-loop effective apparatus action,
describing the long-range quantum corrections, is calculated.
The value of effective gravitational field inferred from the effective
apparatus action is found to be gauge-dependent. A probable
explanation of this result, referring to a peculiarity of
the gravitational interaction, is given.
\end{abstract}


\unitlength=1pt


\section{Introduction}

Quantization of fields, like any other quantization procedure,
contains as one of its inalienable traits an ambiguity
in the choice of a set of fundamental variables in terms of which
transition from deterministic classical theory to statistical quantum theory
is performed. In quantum mechanics this
constitutes most of what is called the operator-ordering problem.
In the quantum theory of fields this appears as the problem of
dependence of observables on the choice of field parametrization
or, in gauge theories, on the choice of gauge-fixing conditions.
Nowhere it is better illustrated than in calculation of the effective fields,
i.e. fields incorporating vacuum polarization effects induced by a given
classical solutions. Due to special structure of Lagrangians of
quantum electrodynamics and Yang-Mills theories, gauge dependence of
the corresponding effective fields occurs only beginning with the
two-loop approximation of the perturbation theory. Instead, in quantum gravity,
dependence of the effective gravitational field on the gauge (and
parametrization) is fully displayed already at the one-loop
level\footnote{Explicit calculations of the one-loop divergences of the
effective action for Einstein gravity and $R^2$-gravity in arbitrary gauge
and parametrization can be found in \cite{kazakov1}.}.

Most generally, the problem under consideration can be stated as
the problem of gauge and parametrization dependence of the effective action.
The latter can be defined either as the sum over all one-particle-irreducible
diagrams \cite{salam}, or as the Legendre transform of the logarithm of
the generating functional of Green functions with respect to the sources
\cite{jona}. The latter definition
brings to light the remarkable analogy between classical equations
of motion and the quantum equations describing dynamics of the mean
fields. It suggests natural interpretation of the effective action
as the quantum substitute for its classical counterpart.
However, the above-mentioned problem lacks direct physical
application of this analogy.

On the other hand, as we pointed out at
the beginning, this "drawback" is an obligatory consequence of the change
in the interpretative framework -- non-invariance with respect to transitions
between different sets of fundamental variables, with the help of which
averaging procedures are established, is in the very statistical
nature of quantum theory. It disappears only in the classical limit.
Although this fact is quite obvious, it should be emphasized
that one does not even have to try to prove it, since it is a direct
consequence of Bohr's {\it correspondence principle} which underlies
quantum theory itself. At this point, we would like to recall
that classical conceptions play crucial role in another important notion
of quantum theory -- the process of measurement.
As emphasized in \cite{heis}, the very idea of acquisition
of some definite reading by a measuring device
is essentially classical. In the light of the gauge dependence problem
eventual classical nature of any process of measurement becomes particularly
important. Namely, it raises the question of whether this problem is
just a matter of the formalism of effective action itself, or
indicates the necessity to include measuring apparatus into the
mathematical description of quantum phenomena explicitly, so as to
make it possible to reformulate the theory of effective quantities
in terms characterizing motion of the apparatus. If the former is true,
then we are left with the S-matrix approach as the only reliable,
though very restrictive, means of deriving physical predictions, while
the opposite would mean that we get the most general quantum description
at our disposal\footnote{This alternative, of course, is exhaustive only
as far as the problem of gauge dependence is concerned. An eventual
solution to this problem may well turn out to be unsatisfactory from
other points of view.}.

The crucial role of the process of
measurement in approaching to the problem of gauge dependence
of the effective fields was first put forward by
Dalvit and Mazzitelli \cite{dalvit}, who showed, in the case of quantum
gravity, that the equations of motion (geodesic equation) of a test
particle in the weak static effective gravitational field of a point mass,
calculated in the one-loop approximation up to leading logarithms, are
independent of the choice of linear gauge fixing the general coordinate
invariance. The essential thing here is that while the quantum interaction
of the test particle with the gravitational field is negligible in evaluation
of the total effective field, it is {\it not} when the
equations of the test particle motion are being determined. It turns out
that in the latter case the gauge-dependent contribution originating
from the graviton-test particle interaction just cancels that corresponding
to the ordinary gauge dependence of the mean gravitational field.
Gauge independence of the equations of motion of the classical apparatus
allows one to define the gauge independent effective gauge field as the
field that enters these equations and couples to the measuring device in the
classical fashion.

This important result raises naturally the following questions:
first of all, is the aforesaid cancellation just a lucky
accident conditioned by the model
chosen and the approximations made in \cite{dalvit}, or
represents a general property of gauge interactions? Second, if the
latter is true, what is the formal mathematical reason underlying the
above-mentioned gauge dependence cancellation?
That this cancellation is not a mere chance in power-counting-renormalizable
theories at least in the low-energy limit is shown quite generally
in \cite{kazakov2}. There, the use has been made of the ordinary
Becchi-Rouet-Stora-Tyutin (BRST) symmetry \cite{brst} of the Faddeev-Popov
action \cite{faddeev}, which is not, however, a {\it quantum} symmetry
in the present case, since it includes transformations of the {\it classical}
matter describing the measuring device. This symmetry is expressed
by Slavnov-type identities for Green functions, obtained in
\cite{kazakov2} and called there effective Slavnov identities.
With the help of the renormalization equation following from
these identities, gauge independence of the effective equations
of device motion is proved in the low-energy limit up to leading logarithms.

The purpose of the present paper is to continue investigation of
these matters in the case of quantum gravity. First of all,
it is important to consider the role of the measuring apparatus
from the field-theoretical point of view, i.e., when the apparatus is
described by some classical fields dynamics of which can be determined
by the limiting procedure of transition from the quantum to the classical
field theory. Second, it is desirable to extend the whole analysis
to arbitrary space-time configurations.
Significance of the effective Slavnov identities is that they allow one
to put the problem of gauge dependence, understood in the sense outlined
above, in the most adequate and technically convenient way. Namely,
they permit to escape the necessity of explicit evaluation of the mean fields,
which requires solving the corresponding wave equations and subsequent
substitution of the results into the effective equations of the apparatus
motion, in order to verify the gauge dependence cancellation, -- the way
followed in \cite{dalvit}. Indeed, as shown in \cite{kazakov2},
the problem of calculation of the total gauge dependence of the effective
device action reduces to evaluation of the gauge dependence of
{\it connected} Green functions containing vertices of the
gauge field-device interaction. It is this simplification that allows us to
investigate the problem under consideration in arbitrary gravitational fields.
This is the subject of the present article.
We take as an example the quantized gravitational field measured by
a classical scalar field. As in \cite{dalvit}, the measuring apparatus
is considered as testing (i.e. neglecting its contribution to the total
effective gravitational field), so the results are trivially extended
to the case when the apparatus is described by an arbitrary number
of scalar particles possessing, in particular, internal symmetries
(e.g., the pions), the only condition being one and the same structure of
interaction with the gravitational field for all particles.

A very important peculiarity of the gravitational interaction must
be mentioned here. As was noted in \cite{donoghue},
the notion of classical matter looses its usual meaning in the case
of gravity, simply because the strength of the gravitational interaction
is proportional to the particle mass. As a result, the relative
quantum corrections to the equations of motion of the particle
do not disappear in the limit $m \to \infty.$
In particular, the whole calculation of the gauge dependence of the
effective apparatus action is divided into two large parts:
evaluation of the total (in the sense of \cite{dalvit}) gauge dependence
of the scalar field effective action when the quantum propagation
of the scalar field is neglected (i.e., when its initial action is taken
in the usual classical form), and calculation of the gauge dependence
of the off-mass-shell scalar field form factors in the limit $m \to \infty.$

The first part of this program is carried out in this paper.

Our paper is organized as follows.
We begin in Sec.~\ref{prelim} with a more detailed formulation of the
gauge dependence problem as well as of the approach to it, adopted
in this paper. The effective Slavnov identities for
the generating functionals of Green functions, introduced in
Sec.~\ref{generfunc}, are derived in Sec.~\ref{slavid}.
In essential, this derivation reproduces that given in \cite{kazakov2}.
In Sec.~\ref{logs}, these identities are used in evaluation of the
logarithmic contribution to the $\xi$-dependent part of
the generating functional of connected Green functions, $\xi$ being
the gauge parameter weighting Lorentz-type gauge condition
$a \partial^{\mu} h_{\mu\nu}
+ b \eta^{\alpha\beta}\partial_{\nu} h_{\alpha\beta} = 0$.
We find that, unlike the case of the point-like measuring apparatus
considered in \cite{dalvit}, this contribution is not zero.
The results obtained are discussed in Sec.~\ref{conclud}.

We use the highly condensed notations of DeWitt \cite{dewitt1} throughout
this paper. Also left derivatives with respect to anticommuting variables
are used. The dimensional regularization of all divergent quantities
is supposed.

\section{Preliminaries}\label{prelim}

Before we proceed to calculations, we would like to
give a somewhat more detailed account of the main aspects of the gauge
dependence problem, briefly mentioned in the Introduction.

\subsection{The origin of the problem}\label{origin}

First of all, we would like to point out the close connection
of the gauge dependence problem with the analogous problem
of dependence of the effective action on the choice of field parametrization.
Roughly speaking, imposition of the gauge conditions in a gauge theory,
being equivalent to picking a subset out of the total set of
variables describing the theory, includes, in particular, freedom to
perform arbitrary substitutions among the variables of the subset.
The parametrization dependence problem is just the reflection
of this freedom possessed by any gauge as well as non-gauge field theory.
As a matter of fact, any like the above reasoning has restricted validity
in field theory because of the fundamental locality requirements.
Nevertheless, it is a general result of the Batalin-Vilkovisky method
\cite{batvil} that the behavior of the effective action under variations
of the gauge conditions is essentially the same as under arbitrary
changes of parametrization -- both are represented by anticanonical
transformations (in the sense of the Batalin-Vilkovisky antibracket).
This result is established in full generality in \cite{lavtut} and,
furthermore, is valid in renormalized as well as unrenormalized theory.
In view of this fact, one can speak about either gauge or
parametrization dependence of the effective action.

Let us now turn to the problem itself. As we have already mentioned
above, its origin is in the inevitable ambiguity in the
choice of a set of fundamental variables in terms of which
the quantization procedure is carried out.
Consider a simple example. Let the system be described by an
action $S(\varphi)$ which is a function of a single scalar field
$\varphi(x)$ and suppose that the functional integral measure
can be chosen simply as the product of $d\varphi(x)$. Then
the mean field $\varphi(y)$ is
\begin{eqnarray}\label{exmpl}&&
\langle\varphi(y)\rangle_{\varphi} = {\displaystyle\int} \prod\limits_x d\varphi(x) \varphi(y)\exp\{i S(\varphi) \},
\end{eqnarray}
The subscript $\varphi$ indicates that the role of fundamental dynamical
variable is played here by the field $\varphi$ itself.
Nothing prevents us, however, from taking as fundamental any
other field $\varphi^* = f[\varphi],$ $f[\varphi]$ being an
arbitrary (local) non-degenerate function, in which case Eq.~(\ref{exmpl})
is replaced by
\begin{eqnarray}\label{exmpl1}&&
\langle\varphi^*(y)\rangle_{\varphi^*} =
{\displaystyle\int} \prod\limits_x d\varphi^*(x)
\varphi^*(y) \exp\{i S^*(\varphi^*) \},
\nonumber\\&&
~~S^*(\varphi^*) \equiv S(f^{-1}[\varphi^*]).
\end{eqnarray}
Employing dimensional regularization, so that $\delta(0)$-type expressions
are set equal to zero, we rewrite Eq.~(\ref{exmpl1}) as
\begin{eqnarray}\label{exmpl2}&&
\langle\varphi^*(y)\rangle_{\varphi^*} = {\displaystyle\int}
\prod\limits_x d\varphi(x) f[\varphi(y)]\exp\{i S(\varphi)\}
= \langle f[\varphi(y)]\rangle_{\varphi}.
\end{eqnarray}

Since there is no reason to prefer one way of quantization to the other,
in particular, one definition of the mean field to the other,
one can try at least to compare the two. Obviously, the only way to do this
is to use the relation $\varphi^* = f[\varphi]$. Consider, e.g., the $\varphi$
picture. On the one hand, as it follows from Eq.~(\ref{exmpl}),
the mean field is equal to $\langle \varphi(y)\rangle_{\varphi}.$
On the other hand, one is equally right to take for it the value
$f^{-1}[\langle \varphi^*(y)\rangle_{\varphi^*}]$. Thus, taking into
account Eq.~(\ref{exmpl2}), we conclude that the change
$\varphi^* = f[\varphi]$ of fundamental variables leads to the
following change in the value of the mean field
\begin{eqnarray}\label{exmpl3}
\langle \varphi(y)\rangle_{\varphi} \to
f^{-1}[\langle f[\varphi(y)]\rangle_{\varphi}].
\end{eqnarray}
Conversely, in the $\varphi^*$ picture, one arrives at the rule
\begin{eqnarray}\label{exmpl4}
\langle \varphi^*(y)\rangle_{\varphi^*} \to
f[\langle f^{-1}[\varphi^*(y)]\rangle_{\varphi^*}].
\end{eqnarray}

Assumptions concerning the regularization scheme and the
function $f[\varphi]$, made above, play no role:
the same transformation rule (\ref{exmpl3}) applies in the general case,
as it follows from the results of \cite{ktut}.
Obviously, ambiguity expressed by this rule is an indispensable
consequence of statistical nature of quantum field theory together with
complete equivalence of various pictures (i.e., formulations
in various sets of variables) at the classical level.
We would like to emphasize in this connection that the very use of the
relation $\varphi^* = f[\varphi]$ in comparison of the two quantum
pictures above is by itself essentially classical.
Yet this deprives the transformations (\ref{exmpl3}) of any significance
in the quantum domain, whereas at the classical level these turn,
of course, into a trivial identity. Nevertheless, requirement
of invariance with respect to the transformations (\ref{exmpl3})
is taken in \cite{vil} as the starting point in construction of a
modification of the ordinary effective action, aimed at derivation
of physically sensible predictions. As it follows from what we just
said, apart from the question of whether the action proposed in \cite{vil}
is actually gauge-independent, such an object would have nothing to do with
the genuine problem of gauge dependence. Instead, one is prompted
by the above discussion to seek the way out of this "difficulty" in
reformulation of the theory in classical terms.
This is exactly what is done in \cite{dalvit}. Effective gravitational
field is defined there through the effective equations of motion of a
classical test particle, namely, as the field that takes place of the
classical gravitational field in the ordinary geodesic equation. Obviously,
this is the only definition relevant to the state of affairs in the light
of the process of measurement. Indeed, in any case it is motion of
specific {\it classical} apparatus measuring the field at any
given space-time interval that is only observed. Thus, it
is value of the field entering {\it classical} equations
of motion of {\it the} apparatus that is only important.

Following \cite{dalvit}, we consider the apparatus as {\it testing},
i.e., as an infinitely small disturbance of the gravitational field it
measures. Let us now show how this technical assumption allows one
to simplify the calculations to be performed below.

\subsection{The role of simple connectedness}\label{simpcon}

Since we are interested mainly in answering the question of whether
the effective field, defined in the sense of Dalvit and Mazzitelli,
is actually gauge independent, rather than in its specific value,
especially in the case of quantum gravity, we can simplify evaluation
of the total gauge-dependent part of the effective device action as follows.
As is explained in the Introduction, this part is the sum of two different
contributions. The first is the ordinary explicit gauge dependence of
the effective device action. The second stems from its implicit gauge
dependence through the mean gauge field. It is exactly this dependence
of the mean gravitational field that lacks its physical interpretation.
Let the measuring apparatus be described by a set of classical fields,
denoted as $\phi$. Then, if $\Gamma_{\phi}$ stands for the $\phi$-dependent
part of the generating functional of one-particle-irreducible Green
functions, $\Gamma$, we have for the full variation of the effective
device action under a small change $\delta\xi$ of a gauge parameter $\xi$:
\begin{eqnarray}\label{full}&&
\delta\Gamma_{\phi}(h,\phi,\xi) = \left.\frac{\partial\Gamma_{\phi}}{\partial\xi}\right|_{h,\phi}\delta\xi
+ \left.\frac{\delta\Gamma_{\phi}}{\delta h_{\mu\nu}}\frac{\partial h_{\mu\nu}}{\partial\xi}\right|_{T,\phi}\delta\xi
\equiv \frac{d\Gamma_{\phi}}{d\xi}\delta\xi,
\end{eqnarray}
where $T\equiv \{T^{\mu\nu}\}$ is the source for the gravitational field,
and $h\equiv \{h_{\mu\nu}\}$ is the mean deviation of the metric from
Minkowskian.

Now note, that if the quantity $W_{\phi}$ is defined by analogy with
$\Gamma_{\phi}$, i.e., as the $\phi$-dependent part of the generating
functional of  {\it connected} Green functions, $W$, then
\begin{eqnarray}\label{equiv}
\Gamma_{\phi}(h,\phi,\xi) = \left.W_{\phi}(T,\phi,\xi)\right|_{T \to T(h,\xi)},
\end{eqnarray}
since the device contribution is infinitesimal.

Comparing Eqs. (\ref{full}) and (\ref{equiv}), we arrive at the following
important relation
\begin{eqnarray}\label{main}
\left.\frac{d\Gamma_{\phi}(h,\phi,\xi)}{d\xi}\right|_{h \to h(T,\xi)}
= \frac{\partial W_{\phi}(T,\phi,\xi)}{\partial\xi}.
\end{eqnarray}

Due to this result, transition to the strongly connected diagrams in the
effective Slavnov identities derived below becomes unnecessary.

\section{The quantum action and the generating functionals}\label{generfunc}

The method we approach the gauge dependence problem allows consideration
of arbitrary space-time configurations. The simplest way to set up any desired
is to choose properly the standard source term normally introduced into the
generating functional of Green functions. Thus, the role of the field
$T^{\mu\nu}$ introduced already in Eqs. (\ref{full}--\ref{main}) will
be not only to serve as the variable of the Legendre transformation,
but also to provide a convenient substitute for realistic matter sources.
The only trace of the latter is the "conservation law"
\begin{eqnarray}\label{cons}
\nabla_{\mu} T^{\mu\nu} = 0,
\end{eqnarray}
where the covariant derivatives are defined with respect to the metric $h^{0}$
satisfying
\begin{eqnarray}\label{tree}
\frac{\delta (S + S_{gf})}{\delta h_{\mu\nu}} = - T^{\mu\nu}.
\end{eqnarray}
$S$ is here the action for the gravitational field\footnote{Our notation
is $R_{\mu\nu} \equiv R^{\alpha}_{\mu\alpha\nu} =
\partial_{\alpha}\Gamma^{\alpha}_{\mu\nu} - ...,
~R \equiv R_{\mu\nu} g^{\mu\nu}, ~g\equiv det g_{\mu\nu}, ~g_{\mu\nu} = sgn(+,-,-,-).$
Dynamical variables of the gravitational field
$h_{\mu\nu} = g_{\mu\nu} - \eta_{\mu\nu},
\eta_{\mu\nu} = diag\{+1,-1,-1,-1\}.$}
\begin{eqnarray}&&\label{actionh}
S = - \frac{1}{k^2}{\displaystyle\int} d^4 x \sqrt{-g}R,
\end{eqnarray}
$k$ being the gravitational constant\footnote{We choose units in
which $c = \hbar = k = 1$ from now on.},
while $S_{gf}$ is the gauge-fixing term
\begin{eqnarray}\label{gauge}&&
S_{gf} = \frac{1}{2\xi}\eta^{\alpha\beta} F_{\alpha} F_{\beta},
~~F_{\alpha} = \partial^{\mu} h_{\mu\alpha} - \frac{1 + \beta}{2}\partial_{\alpha} h,
~~h \equiv \eta^{\mu\nu} h_{\mu\nu}.
\end{eqnarray}
$F$ is the most general covariant gauge condition linear in $h$.
Finally, we assume that the measuring device can be described by a
single scalar field $\phi,$ its action being
\begin{eqnarray}&&\label{actionm}
S_{\phi} =  \frac{1}{2}{\displaystyle\int} d^4 x \sqrt{-g}(g^{\mu\nu}\partial_{\mu}\phi \partial_{\nu}\phi - m^2 \phi^2).
\end{eqnarray}
\noindent
The action $S + S_{\phi}$ is invariant under the following
(infinitesimal) gauge transformations\footnote{Indices of the
functions $F, \xi$, as well as of the ghost fields below,
are raised and lowered, if convenient, with the help of Minkowski metric $\eta_{\mu\nu}$.}
\begin{eqnarray}&&\label{gaugesym}
\delta h_{\mu\nu} = \xi^{\alpha}\partial_{\alpha}h_{\mu\nu}
+ (\eta_{\mu\alpha} + h_{\mu\alpha})\partial_{\nu}\xi^{\alpha}
+ (\eta_{\nu\alpha} + h_{\nu\alpha})\partial_{\mu}\xi^{\alpha}
\equiv D_{\mu\nu}^{\alpha}(h)\xi_{\alpha},
\nonumber\\&&
~~\delta\phi = \xi^{\alpha}\partial_{\alpha}\phi \equiv
\tilde{D}^{\alpha}(\phi)\xi_{\alpha},
\end{eqnarray}
where $\xi^{\alpha}$ are the (infinitesimal) gauge functions.
The generators $D,\tilde{D}$ span the closed algebra
\begin{eqnarray}&&\label{algebra}
D_{\mu\nu}^{\alpha,\sigma\lambda} D_{\sigma\lambda}^{\beta}
- D_{\mu\nu}^{\beta,\sigma\lambda} D_{\sigma\lambda}^{\alpha}
= f_{~~~\gamma}^{\alpha\beta} D_{\mu\nu}^{\gamma},
\nonumber\\&&
\tilde{D}^{\alpha}_{1} \tilde{D}^{\beta}
- \tilde{D}^{\beta}_1 \tilde{D}^{\alpha} = f^{\alpha\beta}_{~~~\gamma} \tilde{D}^{\gamma},
\end{eqnarray}
where the "structure constants" $f^{\alpha\beta}_{~~~\gamma}$ are defined by
\begin{eqnarray}&&
f_{~~~\gamma}^{\alpha\beta}\xi_{\alpha}\eta_{\beta} =
\xi_{\alpha}\partial^{\alpha}\eta_{\gamma}
- \eta_{\alpha}\partial^{\alpha}\xi_{\gamma}
\end{eqnarray}

Next, introducing Faddeev-Popov ghost fields
$C_{\alpha}, \bar{C}^{\alpha}$ we write Faddeev-Popov quantum action
\begin{eqnarray}\label{fp}
S_{fp} = S + S_{\phi} + S_{gf} + \bar{C}^{\beta}F_{\beta}^{,\mu\nu}D_{\mu\nu}^{\alpha}C_{\alpha}.
\end{eqnarray}
$S_{fp}$ is still invariant under the following BRST transformations \cite{brst}
\begin{eqnarray}\label{brsta}&&
\delta_{brst}h_{\mu\nu} = D_{\mu\nu}^{\alpha}(h)C_{\alpha}\lambda,
\nonumber\\&&
\delta_{brst}C_{\gamma} = - \frac{1}{2}f^{\alpha\beta}_{~~~\gamma}C_{\alpha}C_{\beta}\lambda,
\nonumber\\&&
\delta_{brst}\bar{C}^{\alpha} = \frac{1}{\xi}F^{\alpha}\lambda,
\end{eqnarray}
\begin{eqnarray}\label{brstb}&&
\delta_{brst}\phi = \tilde{D}^{\alpha}(\phi)C_{\alpha}\lambda,
\end{eqnarray}
$\lambda$ being a constant anticommuting parameter.

BRST transformation rule for the $\phi$-field is here separated from the rest
to emphasize the special role played by the measuring device in the present
formalism. On the one hand, Eqs. (\ref{brsta}), (\ref{brstb}) span
the usual BRST transformation of the quantum action (\ref{fp}).
On the other hand, in derivation of the effective Slavnov identities below,
the $\phi$-field being classical does not take a part
in the {\it quantum} BRST transformation, i.e., transformation of the
path integral measure in the generating functional of Green
functions\footnote{For brevity, the product symbol,
as well as tensor indices of the fields $h_{\mu\nu},
C_{\alpha}, \bar{C}^{\alpha},$ is omitted in the path integral measure.}
\begin{eqnarray}\label{gener}&&
Z[T,\bar{\beta},\beta,K,\tilde{K},L]
= {\displaystyle\int}dh dC d\bar{C} \exp\{i (\Sigma
+ \bar{\beta}^{\alpha}C_{\alpha} + \bar{C}^{\alpha}\beta_{\alpha} + T^{\mu\nu}h_{\mu\nu})\},
\end{eqnarray}
where
\begin{eqnarray}&&
\Sigma = S_{fp}
+ K^{\mu\nu}D_{\mu\nu}^{\alpha}C_{\alpha}
+ \tilde{K}\tilde{D}^{\alpha}C_{\alpha}
+ \frac{1}{2}L^{\gamma} f^{\alpha\beta}_{~~~\gamma}C_{\alpha}C_{\beta},
\nonumber
\end{eqnarray}
$K^{\mu\nu}(x), ~\tilde{K}(x)$ (anticommuting),$ ~L^{\alpha}(x)$(commuting)
being the BRST transformation sources \cite{zinnjustin}.

Before we continue, we would like to make some notes on the form of the
generating functional (\ref{gener}). It is because of its classical nature
the field $\phi$ is absent in the functional integral
measure in Eq.~(\ref{gener}). The quantum propagation of this
field is thereby neglected. Furthermore, the same classical nature
allows $\phi$-field to be considered as a c-function. In terms
of the creation $\hat{a}^{\dagger}$ and annihilation $\hat{a}$ operators this
is expressed as the approximate commutativity
$$\hat{a}\hat{a}^{\dagger} - \hat{a}^{\dagger}\hat{a} \approx 0,$$
justified if the occupation numbers of the quantum states involved
are large compared with the unity. It is worth to recall in this connection
that in quantum field theory the classical requirements can be applied
only to a finite, although otherwise arbitrary, region of field spectrum.
In particular, the above condition on the value of the occupation
numbers inevitably becomes meaningless when applied to all energies.

We see, therefore, that the above form of the generating functional is an
immediate consequence of Bohr's {\it correspondence principle} underlying the
whole quantum theory.

It is to be mentioned, however, that it is in the case of
the gravitational interaction where the usual procedure of transition
to the classical limit does not work. Namely, as we mentioned
in the Introduction, it is senseless to
increase the apparatus mass in order to suppress its quantum
contribution, since the same mass multiplies vertices of the
graviton-apparatus interaction. In short, there is no such thing as
the classical source for gravity \cite{donoghue}. In this paper,
we follow \cite{dalvit}, and put the device action into the classical
form (\ref{actionm}) "by hands".

As we mentioned in Sec.~\ref{origin}, the parametrization dependence problem
is in fact a part of the more general problem of gauge dependence.
Below we consider the latter case as the more familiar. To illustrate
the essence of our approach as well as of the main result it is
sufficient to consider the most important kind of the gauge dependence,
namely, dependence on the weighting parameter $\xi$.
To accomplish this, we modify the quantum action adding the term
\begin{eqnarray}\label{yform}
Y F_{\alpha}\bar{C}^{\alpha},
\nonumber
\end{eqnarray}
$Y$ being a constant anticommuting parameter \cite{nielsen}.
Thus we write the generating functional of Green functions as
\begin{eqnarray}\label{genernew}&&
Z[T,\bar{\beta},\beta,K,\tilde{K},L,Y]
= {\displaystyle\int}dh dC d\bar{C} \exp\{i (\Sigma
\nonumber\\&&
+ Y F_{\alpha}\bar{C}^{\alpha}
+ \bar{\beta}^{\alpha}C_{\alpha} + \bar{C}^{\alpha}\beta_{\alpha} + T^{\mu\nu}h_{\mu\nu})\}.
\end{eqnarray}

To complete our definitions, we introduce the generating functional of
connected Green functions
\begin{eqnarray}\label{defw}
\tilde{W}[T,\bar{\beta},\beta,K,\tilde{K},L,Y]=
- i \ln Z[T,\bar{\beta},\beta,K,\tilde{K},L,Y],
\end{eqnarray}
and then define the effective action $\tilde{\Gamma}$ in the usual way
as the Legendre transform of $\tilde{W}$ with respect to the mean fields
\begin{eqnarray}&&\label{meanh}
h_{\mu\nu} = \frac{\delta \tilde{W}}{\delta T^{\mu\nu}},
\end{eqnarray}
\begin{eqnarray}&&\label{meanc}
~C_{\alpha} = \frac{\delta \tilde{W}}{\delta\bar{\beta}^{\alpha}},
\end{eqnarray}
\begin{eqnarray}&&\label{meanbc}
~\bar{C}^{\alpha} = - \frac{\delta \tilde{W}}{\delta\beta_{\alpha}},
\end{eqnarray}
(denoted by the same symbols as the corresponding field operators):
\begin{eqnarray}&&
\tilde{\Gamma}[h,C,\bar{C},K,\tilde{K},L,Y]
= \tilde{W} [T,\bar{\beta},\beta,K,\tilde{K},L,Y]
\nonumber\\&&
-   \bar{\beta}^{\alpha}C_{\alpha} - \bar{C}^{\alpha}\beta_{\alpha} - T^{\mu\nu}h_{\mu\nu},
\nonumber
\end{eqnarray}

In the standard interpretation, the reciprocal to equations (\ref{meanh}),
\begin{eqnarray}\label{meanrcp}
\frac{\delta\Gamma}{\delta h_{\mu\nu}} = - T^{\mu\nu},
~~\Gamma [h] \equiv \tilde{\Gamma}[h,0,0,...]
\end{eqnarray}
are the effective equations of motion for the full quantum corrected field
$h_{\mu\nu}$.

\section{The effective Slavnov identities}\label{slavid}

In terms of Green functions, the symmetry with respect to the
transformations (\ref{gaugesym}) is expressed by Slavnov-type identities,
to derive which we perform a BRST shift (\ref{brsta}) of integration
variables in the path integral (\ref{genernew}). Unlike the usual case,
however, the modified quantum action $\Sigma$ is not invariant
under this operation, since, as we have mentioned above,
the classical field $\phi$ does not take
a part in it. Therefore, we obtain the following identity
\begin{eqnarray}&&\label{slav}
\hspace{-1cm}
0 = {\displaystyle\int}dh dC d\bar{C}
\left[i \frac{\delta S_{\phi}}{\delta h_{\mu\nu}} D^{\alpha}_{\mu\nu} C_{\alpha}
+ i Y \bar{C}^{\alpha} F_{\alpha}^{,\mu\nu} D^{\beta}_{\mu\nu} C_{\beta}
+ i \frac{Y}{\xi} F_{\alpha}^{2}
- \frac{i}{2}\tilde{K}\tilde{D}^{\gamma} f^{\alpha\beta}_{~~~\gamma} C_{\alpha} C_{\beta}
\right.
\nonumber\\&&
\left.
\hspace{-1cm}
+ T^{\mu\nu} \frac{\delta}{\delta K^{\mu\nu}}
- \bar{\beta}^{\alpha}\frac{\delta}{\delta L^{\alpha}}
- i \beta_{\alpha}\frac{F^{\alpha}}{\xi}
\right]
\exp\{i (\Sigma
+ Y F_{\alpha}\bar{C}^{\alpha}
+ \bar{\beta}^{\alpha}C_{\alpha} + \bar{C}^{\alpha}\beta_{\alpha}
+ T^{\mu\nu}h_{\mu\nu})\}.
\end{eqnarray}
\noindent
Since $S_{\phi}$ is itself invariant under BRST-transformations (\ref{brsta}),
(\ref{brstb}),
\begin{eqnarray}&&
\frac{\delta S_{\phi}}{\delta h_{\mu\nu}} D^{\alpha}_{\mu\nu} C_{\alpha}
= - \frac{\delta S_{\phi}}{\delta\phi}\tilde{D}^{\alpha} C_{\alpha}.
\end{eqnarray}
Using this identity, we rewrite the first term in the square brackets
on the left of Eq.~(\ref{slav}) as follows
\begin{eqnarray}&&
{\displaystyle\int}dh dC d\bar{C} \frac{\delta S_{\phi}}{\delta h_{\mu\nu}} D^{\alpha}_{\mu\nu} C_{\alpha} \exp\{\cdot\cdot\cdot\} =
- {\displaystyle\int}dh dC d\bar{C} \frac{\delta S_{\phi}}{\delta\phi} \tilde{D}^{\alpha} C_{\alpha} \exp\{\cdot\cdot\cdot\}
\nonumber\\&&
= - \frac{1}{i}\frac{\delta}{\delta\phi}{\displaystyle\int}dh dC d\bar{C} \tilde{D}^{\alpha} C_{\alpha} \exp\{\cdot\cdot\cdot\} +
{\displaystyle\int}dh dC d\bar{C} \tilde{D}^{\alpha} C_{\alpha} \tilde{K}\frac{\delta\tilde{D}^{\beta}}{\delta\phi}C_{\beta} \exp\{\cdot\cdot\cdot\}
\nonumber\\&&
= \frac{\delta^{2} Z}{\delta\phi\delta\tilde{K}} + {\displaystyle\int}dh dC d\bar{C} \tilde{K}\tilde{D}^{\gamma}
\frac{1}{2}f_{~~~\gamma}^{\alpha\beta} C_{\alpha} C_{\beta} \exp\{\cdot\cdot\cdot\},
\end{eqnarray}

\noindent
where the second of Eqs.~(\ref{algebra}), and the locality of
generator $\tilde{D}(\phi)$ together with the property
$\delta (0) = 0$ were taken into account.

Next, the second term in the square brackets in Eq.~(\ref{slav})
can be transformed with the help of the quantum ghost equation of motion,
obtained by performing a shift $\bar{C} \to \bar{C} + \delta\bar{C}$
of integration variables in the functional integral (\ref{gener}):

\begin{eqnarray}\label{ghosteq}&&
{\displaystyle\int}dh dC d\bar{C}
\left[F_{\gamma}^{,\mu\nu}D_{\mu\nu}^{\alpha}C_{\alpha}
- Y F_{\gamma} + \beta_{\gamma} \right]
\nonumber\\&&
\times\exp\{i (\Sigma
+ Y F_{\alpha}\bar{C}^{\alpha}
+ \bar{\beta}^{\alpha}C_{\alpha} + \bar{C}^{\alpha}\beta_{\alpha}
+ T^{\mu\nu}h_{\mu\nu})\}
= 0,
\nonumber
\end{eqnarray}

\noindent
from which it follows that
\begin{eqnarray}\label{ghosteq1}&&
Y {\displaystyle\int}dh dC d\bar{C}
\left[i \bar{C}^{\gamma}F_{\gamma}^{,\mu\nu}D_{\mu\nu}^{\alpha}C_{\alpha}
+ \beta_{\gamma}\frac{\delta}{\delta\beta_{\gamma}}\right]
\exp\{\cdot\cdot\cdot\} = 0,
\nonumber
\end{eqnarray}

\noindent
where the use of the property $Y^2 = 0$ has been made, and the expression
$\delta\beta_{\gamma}/\delta\beta_{\gamma} \sim \delta(0)$ is again
omitted.
Putting this all together, we rewrite Eq.~(\ref{slav})

\begin{eqnarray}\label{slav1}
\left( T^{\mu\nu}\frac{\delta}{\delta K^{\mu\nu}}
- \bar{\beta}^{\alpha}\frac{\delta}{\delta L^{\alpha}}
- \frac{1}{\xi} \beta_{\alpha}F^{\alpha,\mu\nu}\frac{\delta}{\delta T^{\mu\nu}}
+ i \frac{\delta^{2}}{\delta\phi\delta\tilde{K}}
- Y \beta_{\gamma}\frac{\delta}{\delta\beta_{\gamma}}
- 2 Y\xi\frac{\partial}{\partial\xi}
\right) Z  = 0.
\end{eqnarray}

This is the {\it effective Slavnov identity} for the
generating functional of Green functions we are looking for.
In the case $L = \beta = \bar{\beta} = 0$ it was obtained in \cite{kazakov2}.
In terms of the generating functional of connected Green functions,
it looks like

\begin{eqnarray}\label{slav2}
T^{\mu\nu}\frac{\delta \tilde{W}}{\delta K^{\mu\nu}}
- \bar{\beta}^{\alpha}\frac{\delta \tilde{W}}{\delta L^{\alpha}}
- \frac{1}{\xi} \beta_{\alpha}F^{\alpha,\mu\nu}\frac{\delta \tilde{W}}{\delta T^{\mu\nu}}
+ i \frac{\delta^{2} \tilde{W}}{\delta\phi\delta\tilde{K}}
- \frac{\delta \tilde{W}}{\delta\phi}\frac{\delta \tilde{W}}{\delta\tilde{K}}
- Y \beta_{\gamma}\frac{\delta \tilde{W}}{\delta\beta_{\gamma}}
= 2 Y\xi\frac{\partial \tilde{W}}{\partial\xi}.
\end{eqnarray}

Appearance of the second derivatives in Eqs.~(\ref{slav1}),
(\ref{slav2}) is to be emphasized. Recall that it is presence of higher derivatives
(with respect to the sources or mean fields) that makes application of
the original formulation of Slavnov identities much less straightforward
than of their later reformulation in terms of the sources of BRST
transformations, given by Zinn-Justin \cite{zinnjustin} and used in
our derivation above.
However, even the latter approach does not prevent appearance of the
second derivatives in the effective Slavnov identities.
Although the relation (\ref{main}) makes this fact unimportant
for our consideration, an alternative choice of the quantum action
in the generating functional (\ref{genernew}) exists which
allows one to overcome it, and to obtain the effective Slavnov
identities for proper vertices in the general case of finite
contribution of the measuring device to the
effective gravitational field. This is done in the Appendix.

\section{Gauge dependence of leading low-energy quantum corrections}
\label{logs}

In this section, we present detailed evaluation of the total
$\xi$-dependence of the logarithmic contribution to the
effective device action in arbitrary gravitational field.
As we have seen, this dependence is given by the functional
$\partial W_{\phi}/\partial\xi,$ which is conveniently calculated
with the help of the effective Slavnov identity (\ref{slav2}).

Setting $L = \beta = \bar{\beta} = 0$
in Eq.~(\ref{slav2}) and extracting its $Y$-dependent part, we have
\begin{eqnarray}\label{slav3}
2 \xi\frac{\partial W}{\partial\xi} =
- T^{\mu\nu}\frac{\delta \overline{W}}{\delta K^{\mu\nu}}
+ \frac{\delta W}{\delta\phi}\frac{\delta \overline{W}}{\delta\tilde{K}}
- i \frac{\delta^{2} \overline{W}}{\delta\phi\delta\tilde{K}},
\end{eqnarray}
where $W, \overline{W}$ are defined by $$\tilde{W} = W + Y \overline{W}, $$ and the sources
$K^{\mu\nu}, \tilde{K}$ are also set zero after differentiation.

Finally, extraction of the $\phi$-dependent part of Eq.~(\ref{slav3})
gives
\begin{eqnarray}\label{slav4}
2 \xi\frac{\partial W_{\phi}}{\partial\xi} =
- T^{\mu\nu}\frac{\delta \overline{W}_{\phi}}{\delta K^{\mu\nu}}
+ \frac{\delta W_{\phi}}{\delta\phi}\frac{\delta \overline{W}_{\bar{\phi}}}{\delta\tilde{K}}
- i \frac{\delta^{2} \overline{W}_{\phi}}{\delta\phi\delta\tilde{K}},
\end{eqnarray}
where $\overline{W}_{\bar{\phi}}$ denotes the $\phi$-independent part of the
functional $\overline{W}$.
Since the source $\tilde{K}$ is set here equal to zero after differentiation,
the notion of $\phi$-dependence retains its meaning given
in Sec.~\ref{simpcon}.

Thus, to determine the total $\xi$-dependence of the effective apparatus
action one has to evaluate the right hand side of Eq.~(\ref{slav4}).
However, because of zero dimensionality of the gravitational field
calculation of infinity of diagrams with arbitrary number of external
graviton lines is required. The standard way round this difficulty is the
use of the background field method together with an appropriate choice
of the gauge conditions \cite{dewitt1}. This leads to the explicitly
gauge invariant effective action \cite{background}, which allows
one to confine the calculation by the lowest order
of the weak field approximation, and then restore
the whole gauge invariant contribution on dimensional grounds.
It should be emphasized, however, that in our case the very use of this
method is under question. Indeed, functional dependence of the gauge
conditions on the background field is to be considered here on equal
footing with the dependence on the gauge parameters $\xi, \beta.$
Therefore, the background field method itself can be used only after
the effective apparatus action is shown to be independent of such
modification of the gauge conditions, just like as in the $S$-matrix
approach the use of this method is justified by the gauge independence
of the scattering matrix.

It is worthwhile to note that exactly
the background field method was used in \cite{dalvit} in evaluating
the quantum corrections to the geodesic equation, as an intermediate
step of which one had to solve the effective equations of motion for the
(background) gravitational field. It is, however, the main achievement of
the method that these equations are gauge invariant \cite{background}.
As was emphasized in \cite{dalvit},
to solve them one is free to choose any gauge conditions.
The solution depends, of course, on this choice, but this is the ordinary
harmless gauge dependence encountered in any classical gauge theory,
which does not lead to gauge dependence of gauge invariant
functionals of this solution.
This, however, raises the question which of the two sets of gauge
conditions -- the one used to fix the gauge freedom of integration
variables in the path integral, or the one used to solve the equations
of motion -- is to be considered as {\it the} gauge fixing for the effective
gravitational field defined in the sense of Dalvit
and Mazzitelli\footnote{In \cite{dalvit}, the two are simply taken to
coincide, putting thereby the question off.}.
It seems that, in the background field method, it is the gauge conditions
imposed on the background field, which are to be considered also as
the gauge-fixing for the effective field, though without detailed proof this
is only a probable conjecture.

It should be emphasized, however, that all these are only particularities
of the background field method itself. We do {\it not} use this method
in our approach, in which the gauge freedom of the mean gravitational field
is automatically fixed\footnote{In the case of singular gauge conditions,
the gauge fixing for the mean field simply coincides with that for the
integration variables. In general, connection between the
two is more complicated.} by the conditions imposed on the path integral
variables, so the above alternative is never raised.
On the other hand, the questions of method are actually irrelevant
to the problem under consideration, since the very classical nature
of the measuring device implies that its effective action {\it is} gauge
invariant. Thus, if the lowest order contribution to the
gauge-dependent part of the effective device action is zero,
then the whole contribution must be zero too.

\begin{figure}
\hspace*{-1cm}
\epsfxsize=7,5cm\epsfbox{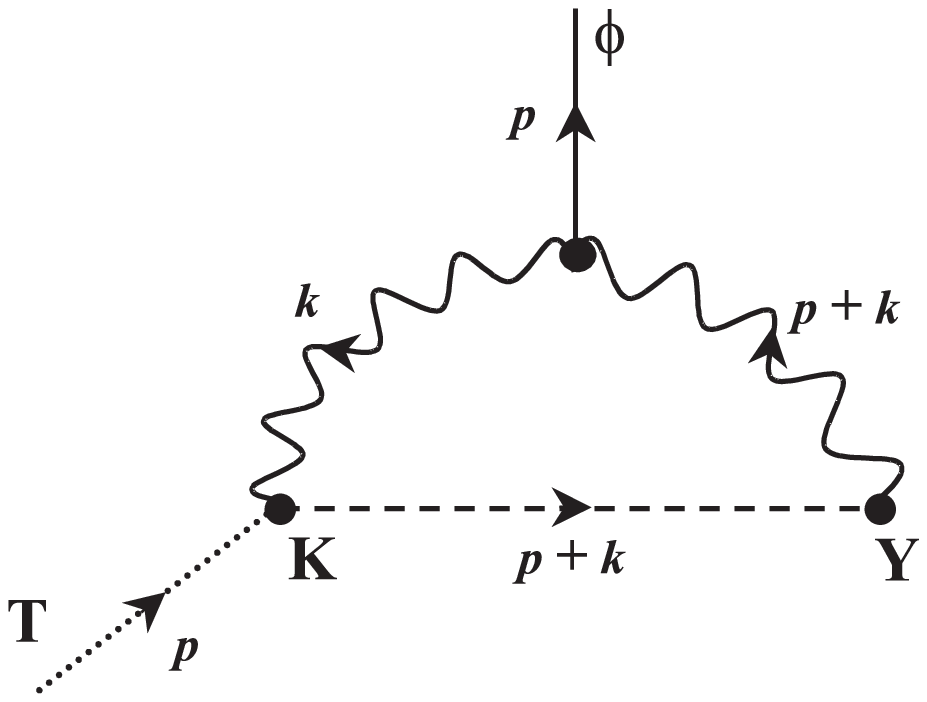}
\put(-115,-5){(a)}
\vskip-4cm
\hspace*{4cm}
\epsfxsize=7,5cm\epsfbox{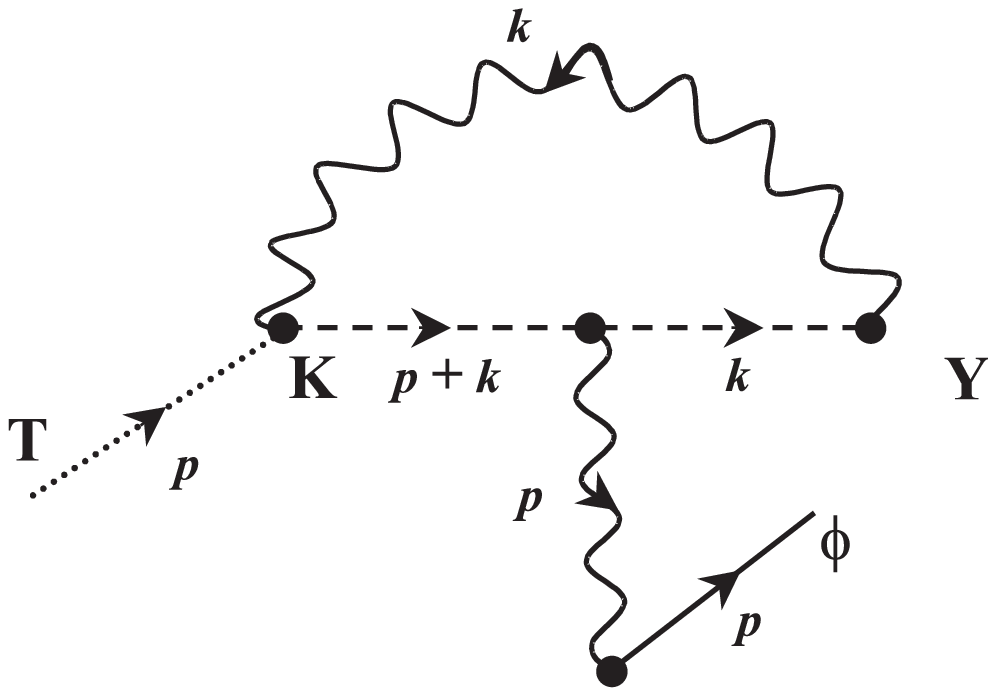}
\put(-115,115){(b)}
\vskip-7,4cm
\hspace*{8,5cm}
\epsfxsize=7,5cm\epsfbox{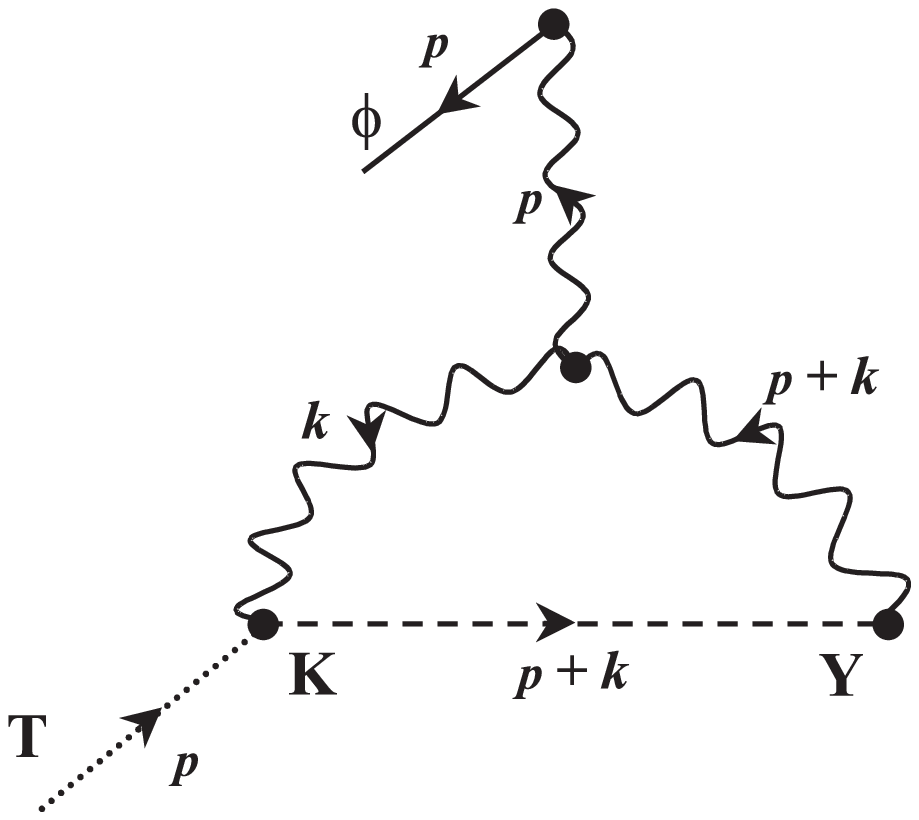}
\put(-75,-5){(c)}
\vskip1,2cm
\hspace*{-1cm}
\epsfxsize=8cm\epsfbox{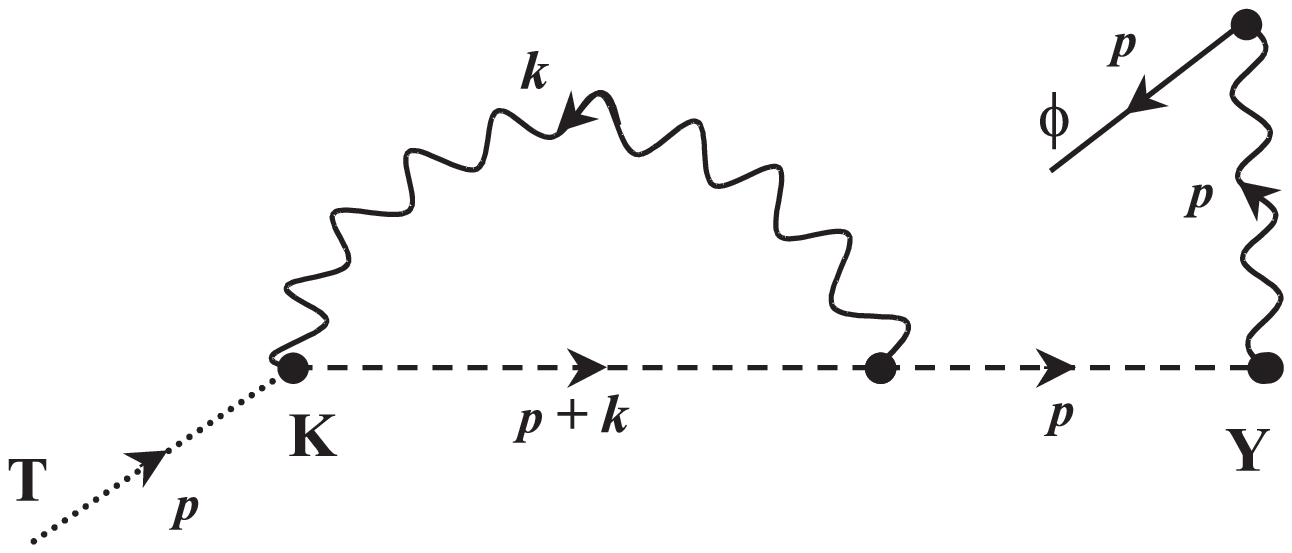}
\put(-95,-5){(d)}
\vskip-3,9cm
\hspace*{9cm}
\epsfxsize=8cm\epsfbox{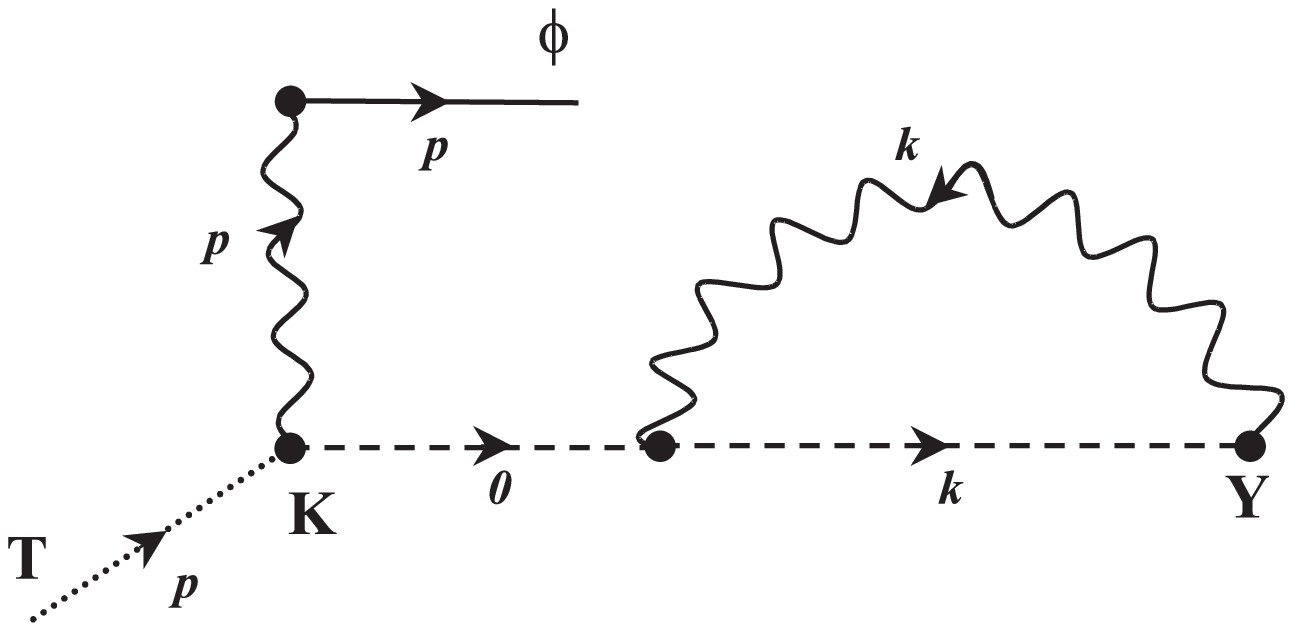}
\put(-115,-5){(e)}
\vskip1,2cm
\hspace*{1cm}
\epsfxsize=14cm\epsfbox{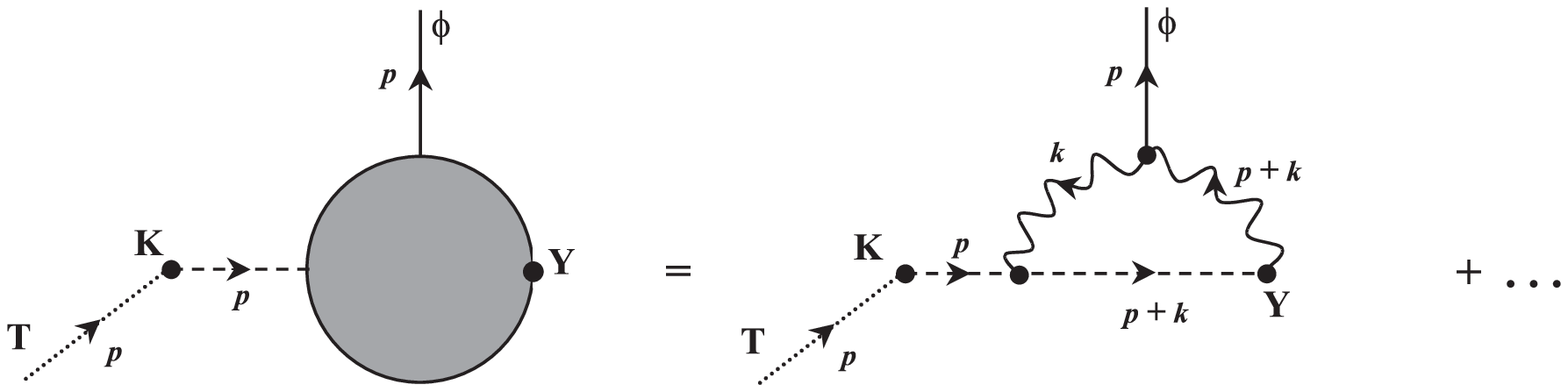}
\put(-245,-5){(f)}
\vskip1cm
\caption{
Diagrams corresponding to the first term in the right hand side
of Eq.~(33). Wavy lines represent gravitons, dashed lines ghosts,
dotted lines the source $T$ for the gravitational field, and
solid lines classical field describing the measuring device. Note that
the latter denote collectively {\it various} functionals of the field
$\phi,$ encountered in Eq.~(33).
As explained in the text, diagrams of the type shown in Fig.~1(f)
do not contribute, so we do not picture them in detail.}
\label{fig1}
\end{figure}

\begin{figure}
\hspace*{-0,1cm}
\epsfxsize=8cm\epsfbox{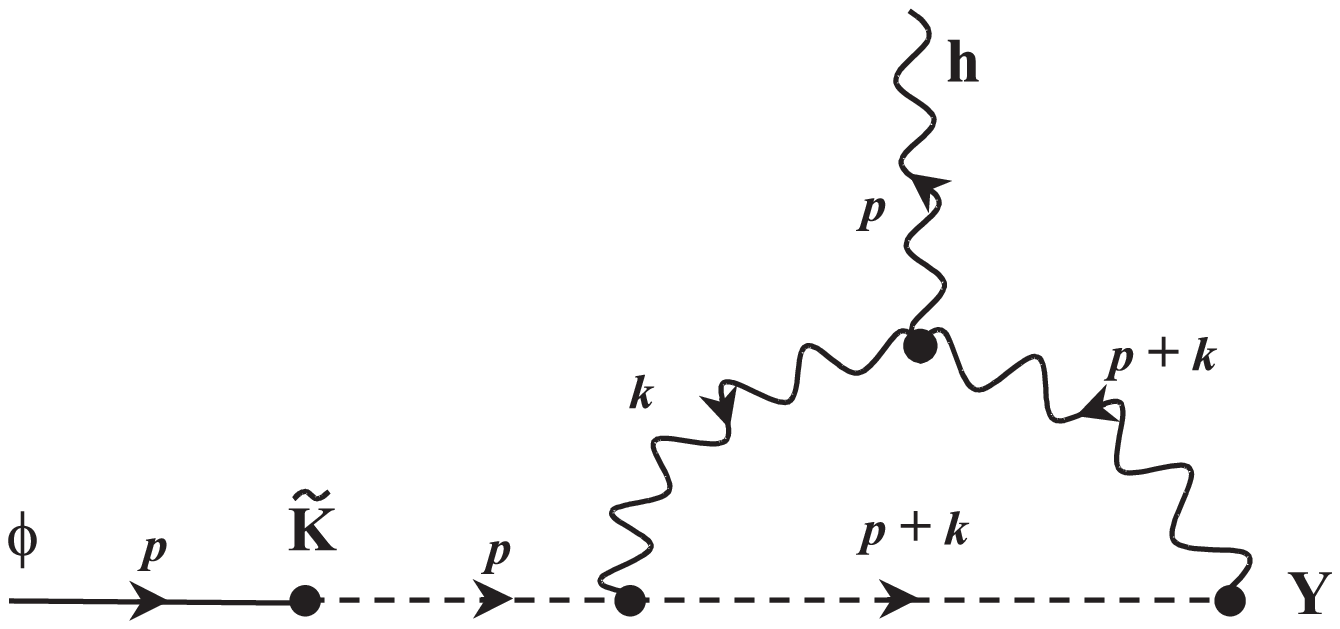}
\put(-135,85){(a)}
\vskip-4,3cm
\hspace*{7,5cm}
\epsfxsize=8cm\epsfbox{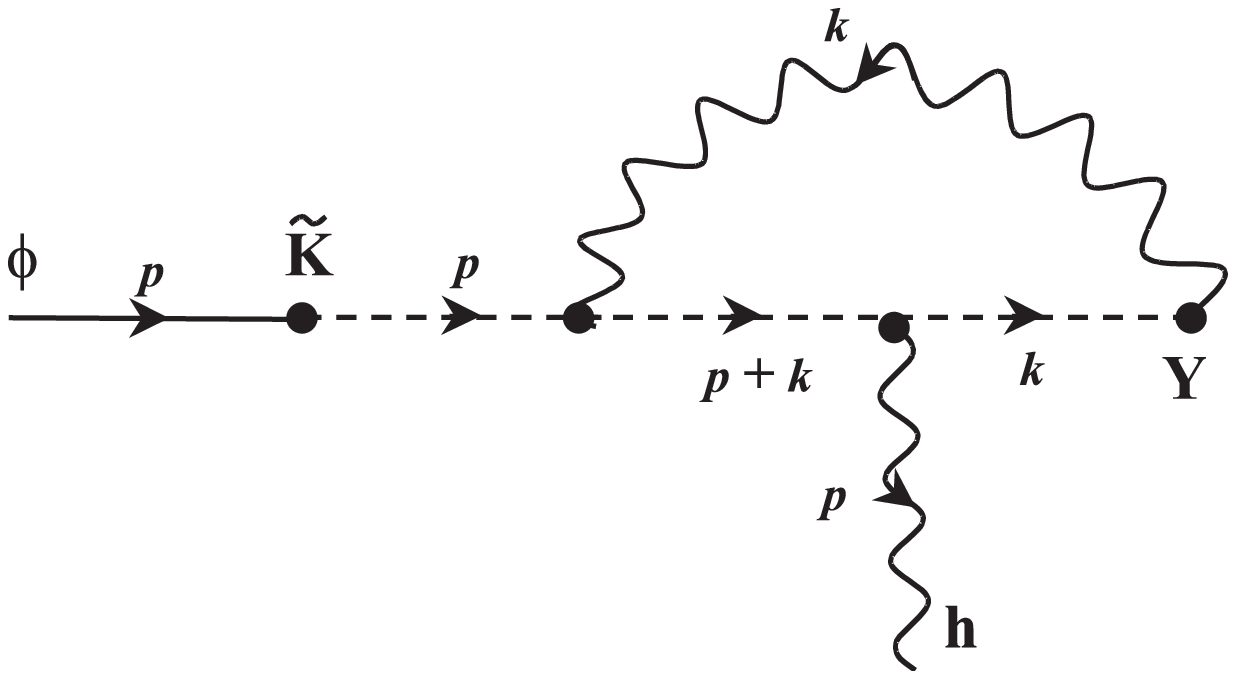}
\put(-85,125){(b)}
\vskip-1cm
\hspace*{-1cm}
\epsfxsize=8cm\epsfbox{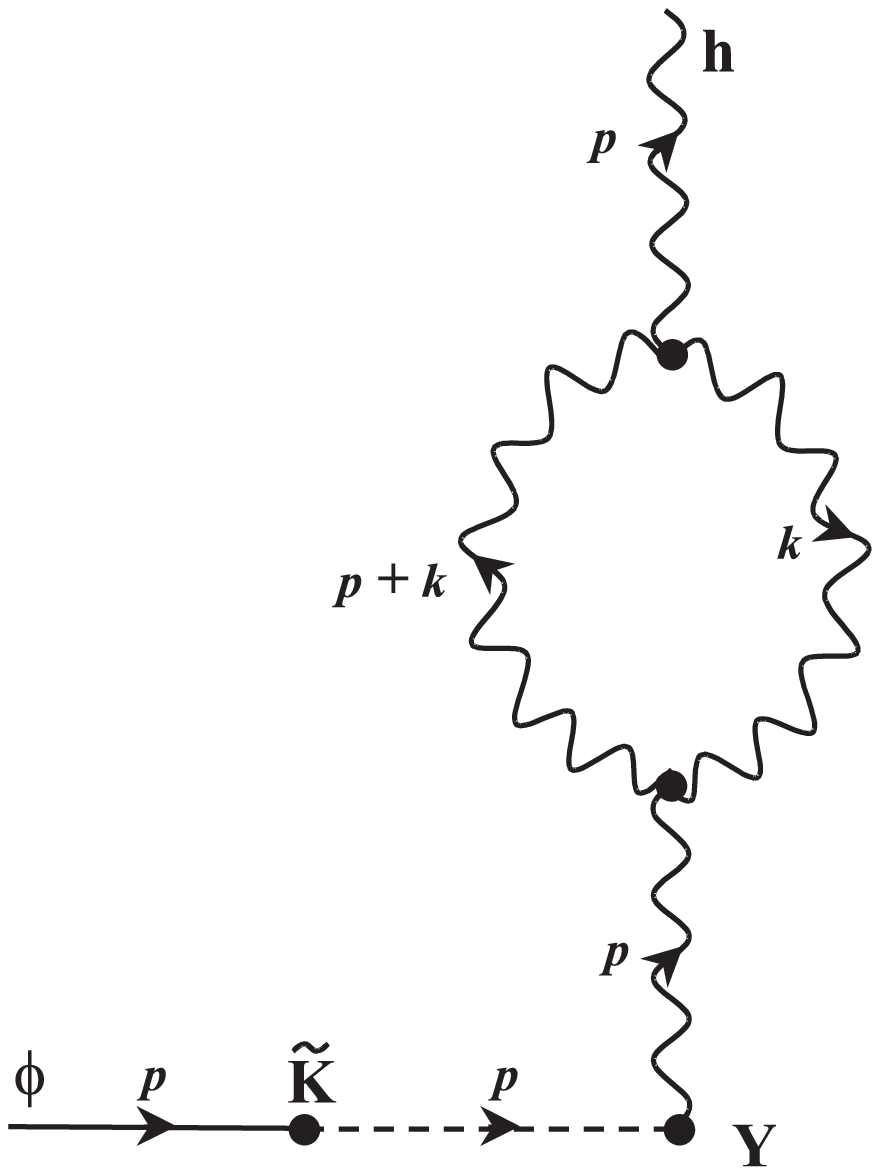}
\put(-165,125){(c)}
\epsfxsize=8cm\epsfbox{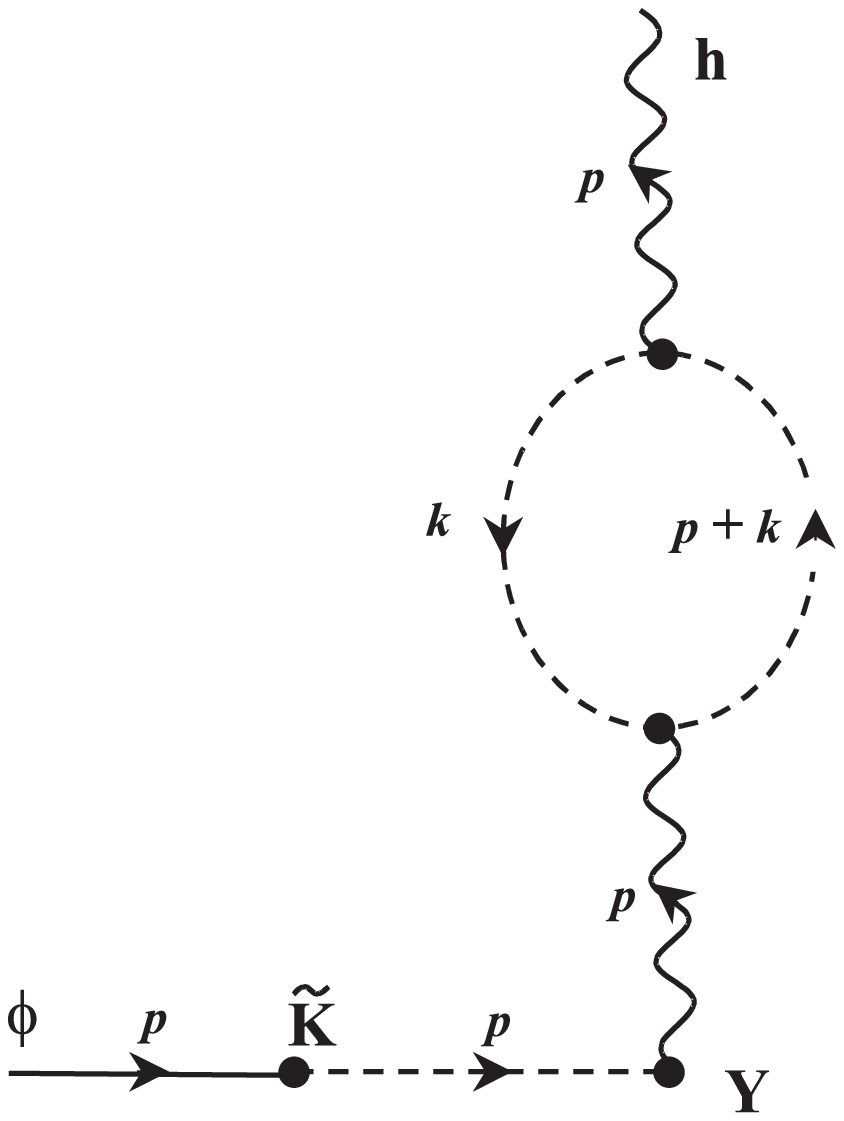}
\put(-165,125){(d)}
\caption{
Diagrams corresponding to the second term in the right hand side
of Eq.~(33).
}
\label{fig2}
\end{figure}

\begin{figure}
\hspace*{-1cm}
\epsfxsize=7,8cm\epsfbox{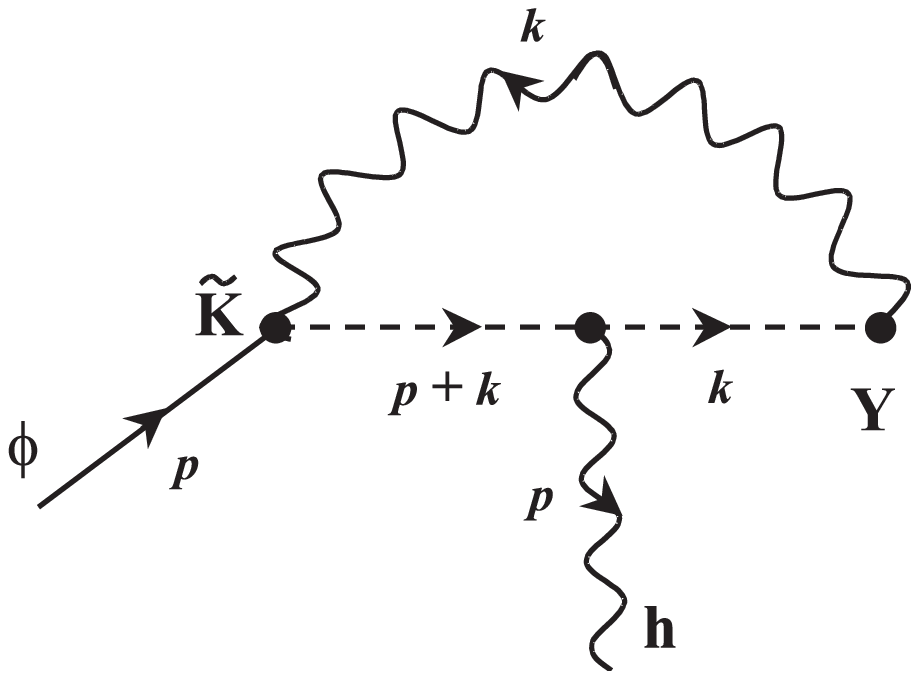}
\put(-120,125){(a)}
\vskip-4,9cm
\hspace*{4,5cm}
\epsfxsize=7,8cm\epsfbox{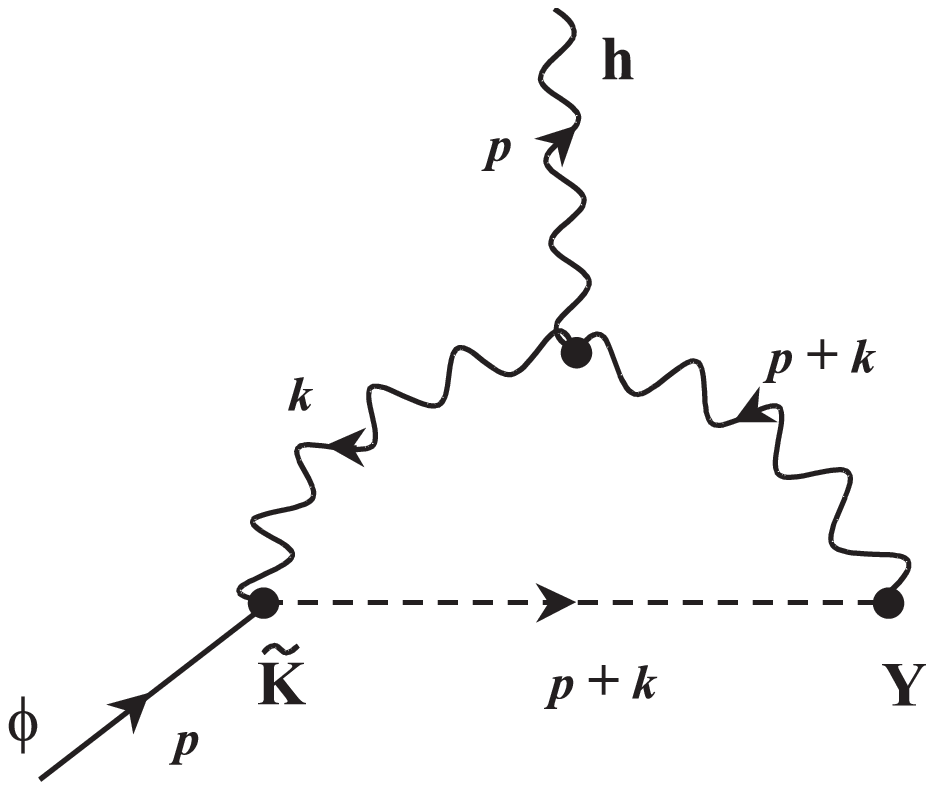}
\put(-120,-15){(b)}
\hspace*{-2,5cm}
\epsfxsize=7,8cm\epsfbox{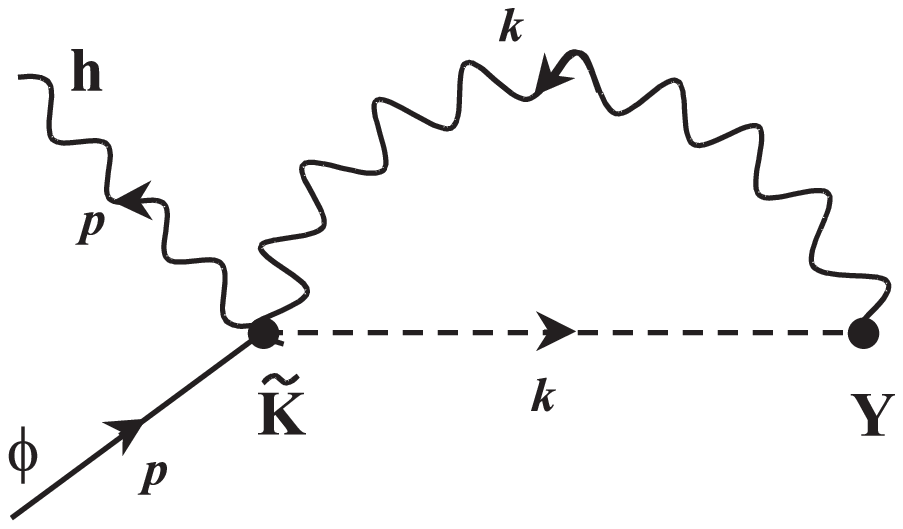}
\put(-120,-15){(c)}
\vskip1cm
\caption{
Diagrams corresponding to the third term in the right hand side
of Eq.~(33). They are obtained from tree diagrams
as shown in Fig.~4.
}
\label{fig3}
\end{figure}

\begin{figure}
\hspace*{-1cm}
\epsfxsize=6cm\epsfbox{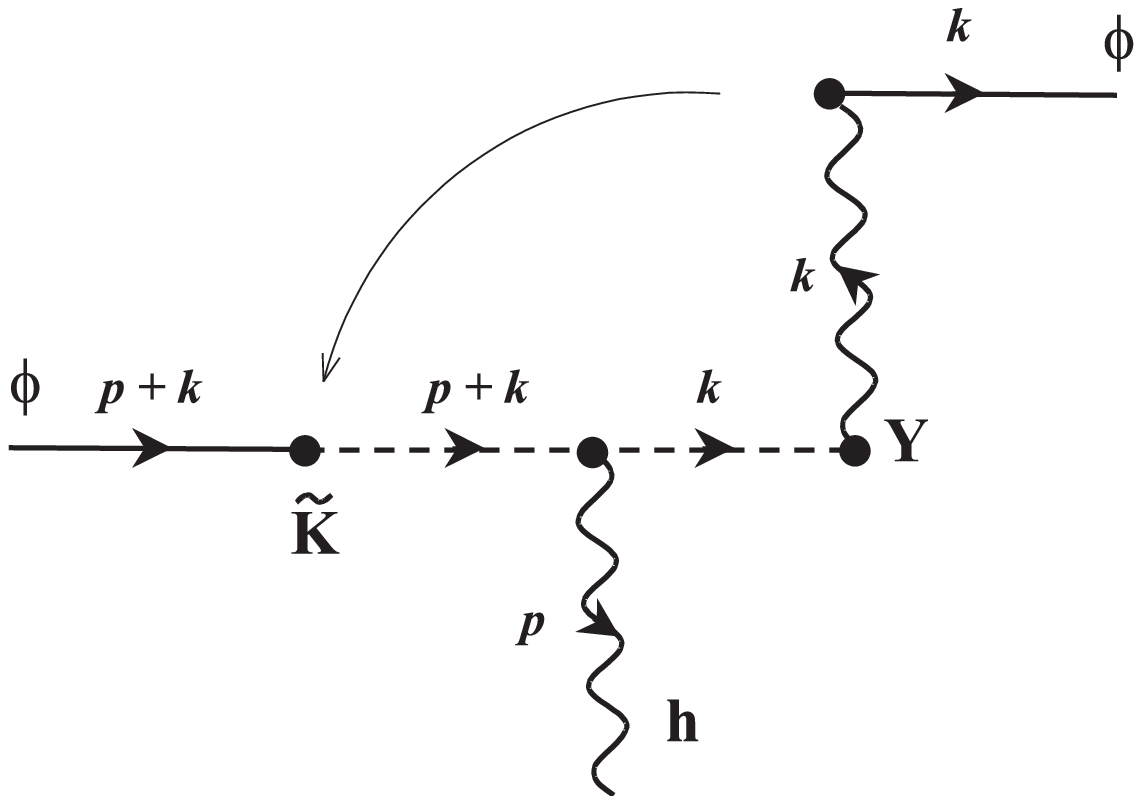}
\put(-140,10){(a)}
\vskip-4,2cm
\hspace*{4,5cm}
\epsfxsize=5cm\epsfbox{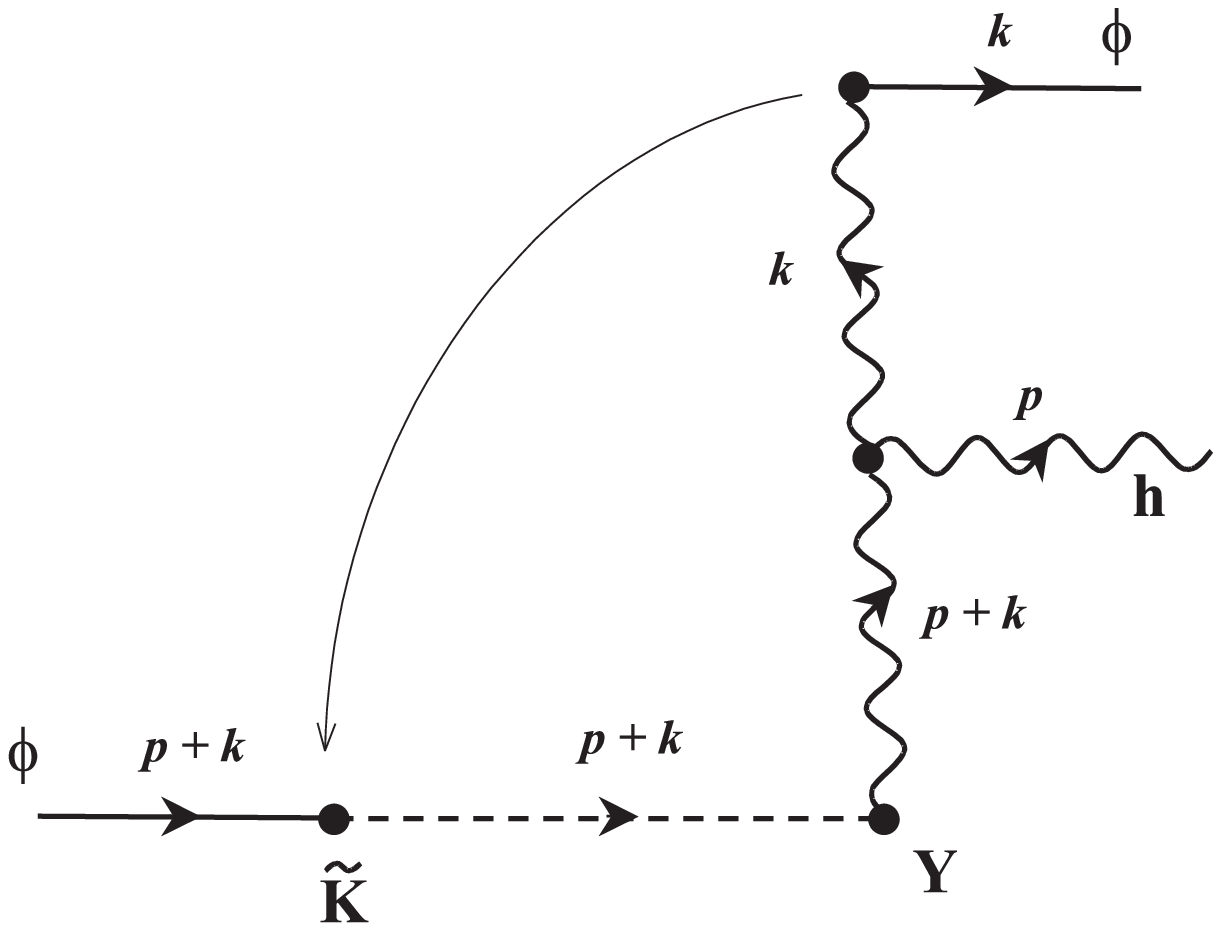}
\put(-80,-25){(b)}
\hspace*{-0,5cm}
\epsfxsize=6cm\epsfbox{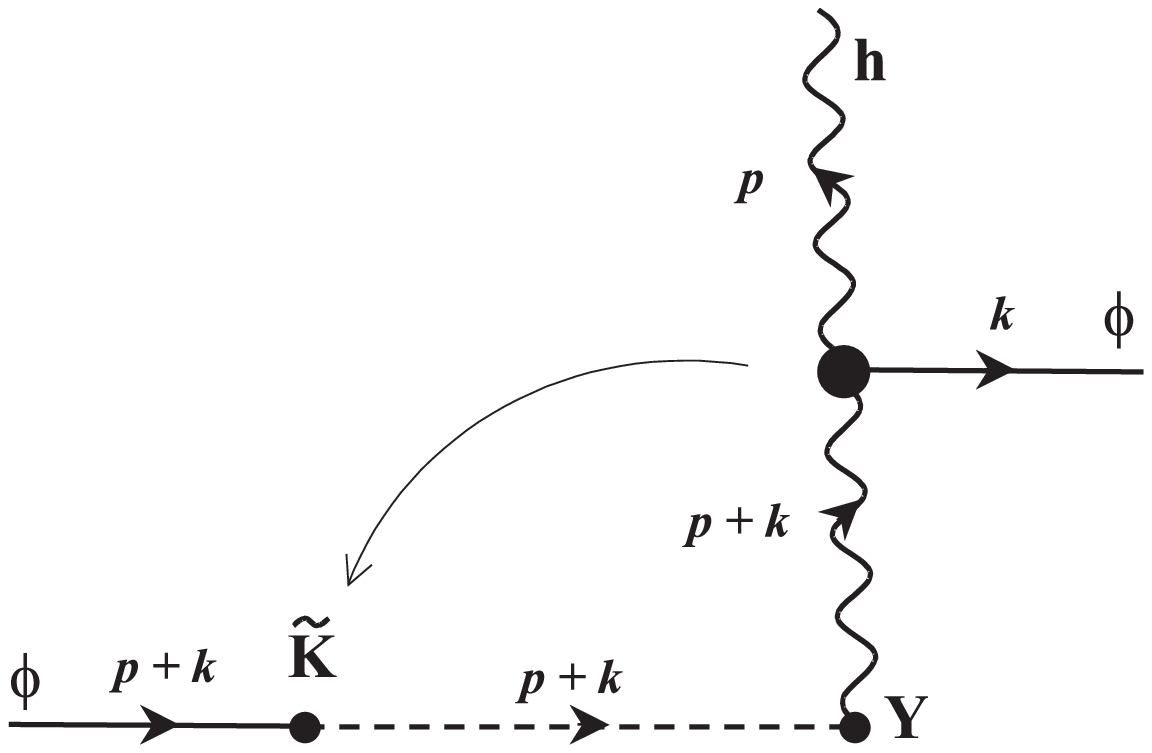}
\put(-80,-25){(c)}
\vskip1cm
\caption{
The tree diagrams from which the one-loop diagrams of
Fig.~3 are formed by confluence of vertices,
shown by the long arrows.}
\label{fig4}
\end{figure}

Let us now proceed to evaluation of the right hand side of Eq.~(\ref{slav4})
in the lowest order of the weak field approximation.
As the subsequent calculation shows, the result is linear in the curvature,
so the lowest non-vanishing contribution is given by diagrams with one
external graviton, pictured in Figs.~\ref{fig1}, \ref{fig2}, and \ref{fig3}
representing the first, second, and third term in the right hand side of
Eq.~(\ref{slav4}), respectively. Note that the diagrams of Fig.\
\ref{fig3}(a,b,c) are obtained from the tree diagrams shown in Fig.\
\ref{fig4}(a,b,c), respectively, by confluence of pairs of vertices,
as shown by the long arrows.

In the first order approximation, Einstein equations read
\begin{eqnarray}&&\label{ein1}
R^{\mu\nu} - \frac{1}{2}\eta^{\mu\nu} R^{\alpha\beta} \eta_{\alpha\beta}
+ T^{\mu\nu} = 0,
\end{eqnarray}
where
\begin{eqnarray}&&
R_{\mu\nu} = \frac{1}{2}(\partial^{\alpha}\partial_{\mu} h_{\alpha\nu}
+ \partial^{\alpha}\partial_{\nu} h_{\alpha\mu}
- \partial^{2} h_{\mu\nu} - \partial_{\mu}\partial_{\nu} h),
\nonumber\\&&
\partial^2 \equiv \eta^{\mu\nu}\partial_{\mu}\partial_{\nu},
~~h \equiv h_{\mu\nu} \eta^{\mu\nu},
\end{eqnarray}
Eq.~(\ref{cons}) becomes the ordinary energy-momentum conservation law
\begin{eqnarray}\label{cons1}
\partial_{\mu} T^{\mu\nu} = 0.
\end{eqnarray}
Since all tensor quantities under consideration ($T^{\mu\nu}, h_{\mu\nu}$ etc.)
are small, their indices are raised and lowered with the help of the
flat-space metric $\eta_{\mu\nu}, \eta^{\mu\nu}.$
Remembering the form of generators of the gauge transformations
Eq.~(\ref{gaugesym}), we see that because of the conservation law
Eq.~(\ref{cons1}), all diagrams of the type shown in Fig.~\ref{fig1}(f) are
equal to zero.

\begin{figure}
\epsfxsize=16cm\epsfbox{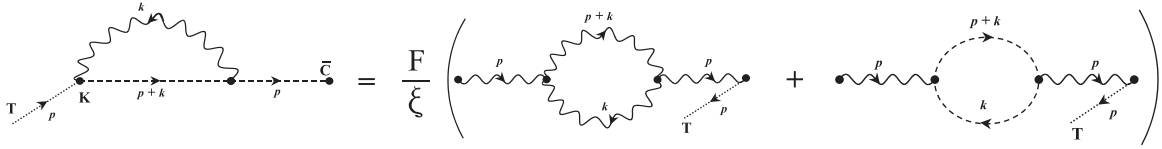}
\vspace{1cm}
\caption{
Graphical representation of the one-loop Slavnov identity (37) contracted
with $T^{\mu\nu}.$}
\label{fig5}
\end{figure}

The ordinary Slavnov identities also help us to show
that the sum of diagrams in Figs.~\ref{fig2}(c) and \ref{fig2}(d)
is equal to the diagram pictured in Fig.~\ref{fig1}(d)
For this purpose, we neglect device contribution,
differentiate Eq.~(\ref{slav2}) twice with respect
to $\beta_{\alpha}$ and $T^{\mu\nu}$, and set all the sources,
{\it including} $T$, equal to zero
\begin{eqnarray}\label{slav5}
\frac{\delta^2 W}{\delta\beta_{\alpha}\delta K^{\mu\nu}}
- \frac{1}{\xi} F^{\alpha,\sigma\lambda}
\frac{\delta^2 W}{\delta T^{\sigma\lambda}\delta T^{\mu\nu}} = 0.
\end{eqnarray}
This is the well-known {\it first} Slavnov identity from which, in particular,
the absence of radiation corrections to the longitudinal part of the
graviton propagator follows. At the tree level, it reads
\begin{eqnarray}\label{slav6}&&
\frac{1}{\xi}F^{\alpha,\mu\nu} G_{\mu\nu\sigma\lambda}(x) =
- D^{(0)\beta}_{\sigma\lambda}\tilde{G}_{\beta}^{\alpha}(x),
\nonumber\\&&
~~D^{(0)\alpha}_{\mu\nu} \equiv D^{\alpha}_{\mu\nu}(h=0),
\end{eqnarray}
and is easily verified with the help of explicit expressions for
the graviton and ghost propagators, $G_{\mu\nu\sigma\lambda},$
$\tilde{G}_{\alpha}^{\beta}$, given below
[see Eqs.~(\ref{g1}), (\ref{gt1})].
At the one-loop approximation, identity (\ref{slav5}), contracted with
$T^{\mu\nu},$ is shown in Fig.~\ref{fig5}.

This figure, together with the identity (\ref{slav6}) and the relation
\begin{eqnarray}&&
\frac{\delta S_{\phi}}{\delta h_{\mu\nu}} D^{\alpha}_{\mu\nu} C_{\alpha}
= - \frac{\delta S_{\phi}}{\delta\phi}\tilde{D}^{\alpha} C_{\alpha},
\end{eqnarray}
makes it clear that the sum of diagrams in Figs.~\ref{fig2}(c,d)
is equal to that pictured in Fig.~\ref{fig1}(d). We would like
to emphasize that the diagrams \ref{fig2}(c,d) are not equal to zero, as
would be the case if the background field method were used.

Finally, diagrams \ref{fig1}(e) and \ref{fig3}(c) are zeros identically.
As for the latter, this follows simply from the fact that there is no external
momentum flow in the dimensionally-regularized loop integral, while in the
former we have $0/0$-type indefiniteness. It is easy to see, however,
that this tadpole is to be set zero. Indeed, let the constant anticommuting
source $Y$ be considered as the limit of a sequence of functions
with infinitely expanding carriers on which the functions have the same
constant value except for immediate neighborhood of the boundaries where
they fall off to zero. Then the above indefiniteness turns into
$q^3/q^2,$ where $q$ is the tending to zero momentum flow
through the $Y$-vertex.

Thus, the diagrams of Figs.~\ref{fig1}(a,b,c,d), \ref{fig2}(a,b),
and \ref{fig3}(a,b) remain to be calculated.

Since each diagram contains only one $\phi$-vertex, two sets of diagrams
corresponding to the two terms in Eq.~(\ref{actionm}) must cancel
independently. Below we present detailed calculation of these diagrams
with $\phi$-vertices generated by the mass term. Evidently, one has
to expand this term up to the second order in the gravitational field
\begin{eqnarray}&&\label{m2phi}
- \frac{1}{2} m^2 \phi^2 \sqrt{- g}
= - \frac{1}{2} m^2 \phi^2
\left(1 +\frac{1}{2} h
+ \frac{1}{8} h^2  - \frac{1}{4} h^{\alpha\beta} h_{\alpha\beta}
+ O(h^3)\right).
\nonumber
\end{eqnarray}
We also calculate:

the second variation of Einstein action (\ref{actionh})

\begin{eqnarray}&&
\hspace{-1cm}
\left.\frac{\delta^2 S}{\delta h_{\sigma\lambda}(x) \delta h_{\mu\nu}(y)}
\right|_{h = 0} =
\left\{- \frac{1}{4}(\eta^{\mu\sigma}\eta^{\nu\lambda}
+ \eta^{\mu\lambda}\eta^{\nu\sigma} - 2 \eta^{\mu\nu}\eta^{\sigma\lambda})
\partial^2
- \frac{1}{2} \left(\eta^{\sigma\lambda}\partial^\mu \partial^\nu
+ \eta^{\mu\nu}\partial^\sigma \partial^\lambda\right)
\right.
\nonumber\\&&
\left.
\hspace{-1cm}
+ \frac{1}{4}\left(\eta^{\sigma\mu} \partial^\lambda \partial^\nu
+ \eta^{\lambda\mu} \partial^\sigma \partial^\nu
+ \eta^{\sigma\nu} \partial^\lambda \partial^\mu
+ \eta^{\lambda\nu} \partial^\sigma \partial^\mu
\right)\right\}\delta(x - y),
~~\partial^2 \equiv \eta^{\mu\nu}\partial_{\mu}\partial_{\nu},
~~\partial_{\mu} \equiv \frac{\partial}{\partial x^{\mu}},
\nonumber
\end{eqnarray}
\noindent
the graviton propagator $G_{\mu\nu\sigma\lambda}$ defined by
$$\frac{\delta^2 S}{\delta h_{\rho\tau}\delta h_{\mu\nu}}
G_{\mu\nu\sigma\lambda} = - \delta_{\sigma\lambda}^{\rho\tau},$$
\begin{eqnarray}&&\label{g1}
G_{\mu\nu\sigma\lambda} = (\eta_{\mu\sigma}\eta_{\nu\lambda} +
\eta_{\mu\lambda}\eta_{\nu\sigma}
- \eta_{\mu\nu} \eta_{\sigma\lambda})
\frac{1}{\partial^2}
+ 2 (\alpha + 1) (\eta_{\mu\nu} \partial_{\sigma} \partial_{\lambda}
+ \eta_{\sigma\lambda} \partial_{\mu} \partial_{\nu})\frac{1}{\partial^4}
\nonumber\\&&
+ (\xi - 1) (\eta_{\mu\sigma} \partial_{\nu} \partial_{\lambda}
+ \eta_{\mu\lambda} \partial_{\nu} \partial_{\sigma}
+ \eta_{\nu\sigma} \partial_{\mu} \partial_{\lambda}
+ \eta_{\nu\lambda} \partial_{\mu} \partial_{\sigma}) \frac{1}{\partial^4}
\nonumber\\&&
+ (4 \alpha^2 \xi - 12 \alpha^2 - 16 \alpha - 4 \xi - 4)
\partial_{\mu} \partial_{\nu} \partial_{\sigma} \partial_{\lambda}
\frac{1}{\partial^6}, ~~\alpha \equiv \frac{1}{\beta - 1},
\end{eqnarray}

\noindent
and the ghost propagator
\begin{eqnarray}&&\label{gt1}
\tilde{G}^{\alpha}_{\beta} = - \frac{\delta^{\alpha}_{\beta}}{\partial^{2}}
- \frac{\beta}{1 - \beta}\frac{\partial^{\alpha}\partial_{\beta}}{\partial^{4}},
\end{eqnarray}
satisfying
\begin{eqnarray}&&
F_{\alpha}^{,\mu\nu}D^{(0)\beta}_{\mu\nu}\tilde{G}^{\gamma}_{\beta}
= - \delta_{\alpha}^{\gamma}.
\nonumber
\end{eqnarray}

The three-graviton vertex encountered in diagrams \ref{fig1}(c),
\ref{fig2}(a), and \ref{fig3}(b), need not be calculated explicitly.
Indeed, with the help of the identity (\ref{slav6}) the graviton propagator
entering the vertex $Y$ can be substituted by the ghost propagator,
the corresponding generator $D^{(0)\alpha}_{\mu\nu}$ being attached to the
three-graviton vertex. The latter, therefore, can be expressed through
the second variation of Einstein action
\begin{eqnarray}&&
\left.\frac{\delta^3 S}{\delta h_{\mu\nu}\delta
h_{\sigma\lambda}\delta h_{\rho\tau}}\right|_{h = 0} D^{(0)\alpha}_{\mu\nu}
+ \left.\frac{\delta^2 S}{\delta h_{\mu\nu}\delta
h_{\rho\tau}}\right|_{h = 0}
\frac{\delta D^{\alpha}_{\mu\nu}}{\delta h_{\sigma\lambda}}
+ \left.\frac{\delta^2 S}{\delta h_{\mu\nu}\delta
h_{\sigma\lambda}}\right|_{h = 0}
\frac{\delta D^{\alpha}_{\mu\nu}}{\delta h_{\rho\tau}} = 0.
\nonumber
\end{eqnarray}
This is obtained by double differentiating the basic identity
\begin{eqnarray}&&
\frac{\delta S}{\delta h_{\mu\nu}} D^{\alpha}_{\mu\nu} = 0.
\end{eqnarray}

Thus, diagrams to be calculated take the following analytical form
\begin{eqnarray}&&
1a = - i \left\{-\frac{1}{2} m^2 \phi^2(-p)\right\} E^{\mu\nu}(p) ~\mu^{\varepsilon} {\displaystyle\int} \frac{d^{4-\varepsilon} k}{(2\pi)^4}
\left(\frac{1}{4}\eta^{\tau\rho} \eta^{\sigma\lambda}
- \frac{1}{2}\delta^{\tau\rho,\sigma\lambda} \right)
\nonumber\\&&
\times G_{\tau\rho\chi\theta}(k)
\left\{k_{\alpha} \delta_{\mu\nu}^{\chi\theta} -
\delta_{\mu\alpha}^{\chi\theta} (k_{\nu} + p_{\nu}) -
\delta_{\nu\alpha}^{\chi\theta} (k_{\mu} + p_{\mu})\right\}
\nonumber\\&&
\times \tilde{G}^{\alpha}_{\beta}(p+k) \xi\tilde{G}^{\beta\gamma}(p+k)
\left\{\eta_{\sigma\gamma} (k_{\lambda} + p_{\lambda}) +
\eta_{\lambda\gamma} (k_{\sigma} + p_{\sigma})\right\},
\nonumber
\end{eqnarray}

\begin{eqnarray}&&
1b = - i \left\{-\frac{1}{2} m^2 \phi^2(-p)\right\} E^{\mu\nu}(p) ~\mu^{\varepsilon} {\displaystyle\int} \frac{d^{4-\varepsilon} k}{(2\pi)^4}
~\tilde{G}^{\alpha}_{\beta}(p+k)
\nonumber\\&&
\times \left\{\frac{1+\beta}{2}(p^{\beta} + k^{\beta})\eta^{\tau\rho}
- \frac{1}{2}(p^{\tau} + k^{\tau})\eta^{\beta\rho}
- \frac{1}{2}(p^{\rho} + k^{\rho})\eta^{\beta\tau}\right\}
\nonumber\\&&
\times \left\{- p_{\gamma} \delta_{\tau\rho}^{\chi\theta} -
\delta_{\tau\gamma}^{\chi\theta} k_{\rho} -
\delta_{\rho\gamma}^{\chi\theta} k_{\tau} \right\}
G_{\chi\theta\varphi\psi}(p) \frac{\eta^{\varphi\psi}}{2}
\tilde{G}^{\gamma}_{\delta}(k)\xi\tilde{G}^{\delta\zeta}(k)
\nonumber\\&&
\times \left\{\eta_{\sigma\zeta} k_{\lambda}  +
\eta_{\lambda\zeta} k_{\sigma} \right\}
\left\{k_{\alpha} \delta_{\mu\nu}^{\sigma\lambda} -
\delta_{\mu\alpha}^{\sigma\lambda} (k_{\nu} + p_{\nu}) -
\delta_{\nu\alpha}^{\sigma\lambda} (k_{\mu} + p_{\mu})\right\},
\nonumber
\end{eqnarray}

\begin{eqnarray}&&
1c = - i \left\{-\frac{1}{2} m^2 \phi^2(-p)\right\} E^{\mu\nu}(p) ~\mu^{\varepsilon} {\displaystyle\int} \frac{d^{4-\varepsilon} k}{(2\pi)^4}
~\tilde{G}^{\alpha}_{\beta}(p+k) \xi\tilde{G}^{\beta\gamma}(p+k)
\nonumber\\&&
\times \left(
- S^{,\sigma\lambda~\tau\rho}(p)
\left\{- k_{\gamma} \delta_{\sigma\lambda}^{\chi\theta} +
\delta_{\sigma\gamma}^{\chi\theta} (k_{\lambda} + p_{\lambda}) +
\delta_{\lambda\gamma}^{\chi\theta} (k_{\sigma} + p_{\sigma})\right\}
\right.
\nonumber\\&&
\left.
- S^{,\sigma\lambda~\chi\theta}(k)
\left\{- p_{\gamma} \delta_{\sigma\lambda}^{\tau\rho} +
\delta_{\sigma\gamma}^{\tau\rho} (k_{\lambda} + p_{\lambda}) +
\delta_{\lambda\gamma}^{\tau\rho} (k_{\sigma} + p_{\sigma})\right\}
\right)
\nonumber\\&&
\times G_{\tau\rho\varphi\psi}(k)
G_{\chi\theta\kappa\omega}(p) \frac{\eta^{\kappa\omega}}{2}
\left\{k_{\alpha} \delta_{\mu\nu}^{\varphi\psi} -
\delta_{\mu\alpha}^{\varphi\psi} (k_{\nu} + p_{\nu}) -
\delta_{\nu\alpha}^{\varphi\psi} (k_{\mu} + p_{\mu})\right\},
\nonumber
\end{eqnarray}

\begin{eqnarray}&&
1d = - i \left\{-\frac{1}{2} m^2 \phi^2(-p)\right\} E^{\mu\nu}(p) ~\mu^{\varepsilon} {\displaystyle\int} \frac{d^{4-\varepsilon} k}{(2\pi)^4}
G_{\kappa\omega\chi\theta}(k)
\nonumber\\&&
\times \left\{k_{\alpha} \delta_{\mu\nu}^{\chi\theta} -
\delta_{\mu\alpha}^{\chi\theta} (k_{\nu} + p_{\nu}) -
\delta_{\nu\alpha}^{\chi\theta} (k_{\mu} + p_{\mu})\right\}
\tilde{G}^{\alpha}_{\beta}(p+k)
\nonumber\\&&
\times \left\{\frac{1+\beta}{2}(p^{\beta} + k^{\beta})\eta^{\tau\rho}
- \frac{1}{2}(p^{\tau} + k^{\tau})\eta^{\beta\rho}
- \frac{1}{2}(p^{\rho} + k^{\rho})\eta^{\beta\tau}\right\}
\nonumber\\&&
\times \left\{- k_{\gamma} \delta_{\tau\rho}^{\kappa\omega} -
\delta_{\tau\gamma}^{\kappa\omega} p_{\rho} -
\delta_{\rho\gamma}^{\kappa\omega} p_{\tau}\right\}
\tilde{G}^{\gamma}_{\delta}(p) \xi \tilde{G}^{\delta\zeta}(p)
\left\{\eta_{\sigma\zeta} p_{\lambda} +
\eta_{\lambda\zeta} p_{\sigma}\right\}\frac{\eta^{\sigma\lambda}}{2},
\nonumber
\end{eqnarray}

\begin{eqnarray}&&
2a = - i \left\{-\frac{1}{2} m^2 \phi^2(p)\right\} ~\mu^{\varepsilon} {\displaystyle\int} \frac{d^{4-\varepsilon} k}{(2\pi)^4}
~\tilde{G}^{\alpha}_{\beta}(p+k) \xi\tilde{G}^{\beta\gamma}(p+k)
\nonumber\\&&
\times \left(
- S^{,\sigma\lambda~\tau\rho}(p)
\left\{- k_{\gamma} \delta_{\sigma\lambda}^{\chi\theta} +
\delta_{\sigma\gamma}^{\chi\theta} (k_{\lambda} + p_{\lambda}) +
\delta_{\lambda\gamma}^{\chi\theta} (k_{\sigma} + p_{\sigma})\right\}
\right.
\nonumber\\&&
\left.
- S^{,\sigma\lambda~\chi\theta}(k)
\left\{- p_{\gamma} \delta_{\sigma\lambda}^{\tau\rho} +
\delta_{\sigma\gamma}^{\tau\rho} (k_{\lambda} + p_{\lambda}) +
\delta_{\lambda\gamma}^{\tau\rho} (k_{\sigma} + p_{\sigma})\right\}
\right)
\nonumber\\&&
\times h_{\chi\theta}(-p) G_{\tau\rho\varphi\psi}(k)
\left\{\frac{1+\beta}{2} p^{\delta}\eta^{\mu\nu}
- \frac{1}{2}p^{\mu}\eta^{\delta\nu}
- \frac{1}{2}p^{\nu}\eta^{\delta\mu}\right\}
\nonumber\\&&
\times\left\{k_{\alpha} \delta_{\mu\nu}^{\varphi\psi} -
\delta_{\mu\alpha}^{\varphi\psi} (k_{\nu} + p_{\nu}) -
\delta_{\nu\alpha}^{\varphi\psi} (k_{\mu} + p_{\mu})\right\}
\tilde{G}^{\zeta}_{\delta}(p) p_{\zeta},
\nonumber
\end{eqnarray}

\begin{eqnarray}&&
2b = - i \left\{-\frac{1}{2} m^2 \phi^2(p)\right\} ~\mu^{\varepsilon} {\displaystyle\int} \frac{d^{4-\varepsilon} k}{(2\pi)^4}
~\tilde{G}^{\alpha}_{\beta}(p+k)
\nonumber\\&&
\times \left\{\frac{1+\beta}{2}(p^{\beta} + k^{\beta})\eta^{\tau\rho}
- \frac{1}{2}(p^{\tau} + k^{\tau})\eta^{\beta\rho}
- \frac{1}{2}(p^{\rho} + k^{\rho})\eta^{\beta\tau}\right\}
\nonumber\\&&
\times \left\{- p_{\gamma} h_{\tau\rho}(-p) -
h_{\tau\gamma}(-p) k_{\rho} - h_{\rho\gamma}(-p) k_{\tau} \right\}
\tilde{G}^{\gamma}_{\delta}(k)\xi\tilde{G}^{\delta\zeta}(k)
\nonumber\\&&
\times \left\{\eta_{\sigma\zeta} k_{\lambda}  +
\eta_{\lambda\zeta} k_{\sigma} \right\}
\left\{k_{\alpha} \delta_{\mu\nu}^{\sigma\lambda} -
\delta_{\mu\alpha}^{\sigma\lambda} (k_{\nu} + p_{\nu}) -
\delta_{\nu\alpha}^{\sigma\lambda} (k_{\mu} + p_{\mu})\right\}
\nonumber\\&&
\times \left\{\frac{1+\beta}{2} p^{\epsilon}\eta^{\mu\nu}
- \frac{1}{2}p^{\mu}\eta^{\epsilon\nu}
- \frac{1}{2}p^{\nu}\eta^{\epsilon\mu}\right\}
\tilde{G}^{\eta}_{\epsilon}(p) p_{\eta},
\nonumber
\end{eqnarray}

\begin{eqnarray}&&
3a = - i \left\{-\frac{1}{2} m^2 \phi^2(p)\right\} ~\mu^{\varepsilon} {\displaystyle\int} \frac{d^{4-\varepsilon} k}{(2\pi)^4}
~\tilde{G}^{\alpha}_{\beta}(p+k) p_{\alpha}
\left\{\eta_{\sigma\zeta} k_{\lambda}  +
\eta_{\lambda\zeta} k_{\sigma} \right\}\frac{\eta^{\sigma\lambda}}{2}
\nonumber\\&&
\times \left\{\frac{1+\beta}{2}(p^{\beta} + k^{\beta})\eta^{\tau\rho}
- \frac{1}{2}(p^{\tau} + k^{\tau})\eta^{\beta\rho}
- \frac{1}{2}(p^{\rho} + k^{\rho})\eta^{\beta\tau}\right\}
\nonumber\\&&
\times \left\{- p_{\gamma} h_{\tau\rho}(-p) -
h_{\tau\gamma}(-p) k_{\rho} - h_{\rho\gamma}(-p) k_{\tau} \right\}
\tilde{G}^{\gamma}_{\delta}(k)\xi\tilde{G}^{\delta\zeta}(k),
\nonumber
\end{eqnarray}

\begin{eqnarray}&&
3b = - i \left\{-\frac{1}{2} m^2 \phi^2(p)\right\} ~\mu^{\varepsilon} {\displaystyle\int} \frac{d^{4-\varepsilon} k}{(2\pi)^4}
~p_{\alpha} \tilde{G}^{\alpha}_{\beta}(p+k) \xi\tilde{G}^{\beta\gamma}(p+k)
\nonumber\\&&
\times \left(
- S^{,\sigma\lambda~\tau\rho}(p)
\left\{- k_{\gamma} \delta_{\sigma\lambda}^{\chi\theta} +
\delta_{\sigma\gamma}^{\chi\theta} (k_{\lambda} + p_{\lambda}) +
\delta_{\lambda\gamma}^{\chi\theta} (k_{\sigma} + p_{\sigma})\right\}
\right.
\nonumber\\&&
\left.
- S^{,\sigma\lambda~\chi\theta}(k)
\left\{- p_{\gamma} \delta_{\sigma\lambda}^{\tau\rho} +
\delta_{\sigma\gamma}^{\tau\rho} (k_{\lambda} + p_{\lambda}) +
\delta_{\lambda\gamma}^{\tau\rho} (k_{\sigma} + p_{\sigma})\right\}
\right)
\nonumber\\&&
\times h_{\chi\theta}(-p) G_{\tau\rho\mu\nu}(k)\frac{\eta^{\mu\nu}}{2},
\nonumber
\end{eqnarray}

where $E^{\mu\nu}$ stands for the Einstein tensor
\begin{eqnarray}
E^{\mu\nu} = R^{\mu\nu} - \frac{1}{2}\eta^{\mu\nu} R_{\alpha\beta} \eta^{\alpha\beta},
\end{eqnarray}
$\phi^2(p)$ -- Fourier transform of the square of the scalar field,
$\mu$ -- arbitrary mass scale, and $\varepsilon = 4 - d,$ $d$ being the
dimensionality of space-time.

As we see, all Feynman integrals we need to calculate have the following form
\begin{eqnarray}\label{int}&&
I(p) = \mu^{\varepsilon}{\displaystyle\int}d^{4-\varepsilon}k f(p,k),
\end{eqnarray}
where $f(p,k)$ is the product of the graviton and ghost propagators and of
the vertex factors.
Since we neglect quantum propagation of the scalar field,
there is no dimensional parameters in the integrands.
Therefore, as a simple dimensional analysis shows,
$I(p)$ have the following structure
\begin{eqnarray}\label{int1}&&
I = c_1 p^{N} \left(\frac{\mu^2}{p^2}\right)^{\frac{\varepsilon}{2}}
\left[\frac{1}{\varepsilon} + c_2\right]
= c_1 p^{N}
\left[\frac{1}{\varepsilon} - \frac{1}{2}\ln\left(\frac{p^2}{\mu^2}\right)
+ c_2 + O(\varepsilon)\right],
\end{eqnarray}
$c_1, c_2$ and $N$ being some numbers depending on the specific
form of $f(p,k)$. It follows from Eq.~(\ref{int1}) that one can obtain the
logarithmic contribution from divergent one substituting
\begin{eqnarray}\label{alg}&&
\frac{1}{\varepsilon} \to - \frac{1}{2}\ln\left(\frac{p^2}{\mu^2}\right).
\end{eqnarray}
All the above Feynman integrals are ultraviolet divergent.
It is important, on the other hand,
that they are free of infrared divergences.
Indeed, denominators in these integrals are the products of only
two scalars -- $(p + k)^2$ and $k^2.$ If we rewrite also every $p$
entering the vertex factors as $(p + k) - k,$ then the integrands
take the form of sums of products of powers $(p + k)^{n}$ and $k^{m}.$
Since all diagrams are ultraviolet divergent, we have
$n + m \ge - 4.$ On the other hand, infrared divergences appear only
if $n \le - 4,$ or $m \le - 4$, and, therefore, we have $m \ge 0$, or
$n \ge 0,$ respectively. In either case the dimensionally regularized loop integrals
turn into zero.

Thus, to calculate the logarithmic contribution of the diagrams
it is sufficient to find their ultraviolet divergences.
For this purpose, we expand all denominators in a finite series in powers
of $p/k$ keeping only first $N$ terms.
It is convenient to apply identity (\ref{slav6}) to the $Y$-vertex
in all diagrams under consideration, since then such an expansion
is to be performed on the ghost propagator only

\begin{eqnarray}&&
\tilde{G}^{\alpha}_{\beta}(p + k) =
- \frac{\delta^{\alpha}_{\beta}}{k^2}
\left(1 - \frac{2 (p k)}{k^2}
- \frac{p^2}{k^2} + \frac{4 (p k)^2}{k^4} + \frac{4 (p k) p^2}{k^4}
- \frac{8 (p k)^3}{k^6}
\right)
\nonumber\\&&
- \frac{\beta}{1 - \beta}
\frac{(k + p)^{\alpha} (k + p)_{\beta}}{k^4}
\left(1 - \frac{4 (p k)}{k^2} - \frac{2 p^2}{k^2}
+ \frac{12 (p k)^2}{k^4}
+ \frac{12 (p k) p^2}{k^4} - \frac{32 (p k)^3}{k^6}\right)
\nonumber\\&&
+ O\left(\frac{1}{k^6}\right).
\nonumber
\end{eqnarray}

Now, calculation of diagrams is straightforward. The tensor multiplication
as well as integration over angles in the momentum space is
performed with the help of the New Tensor Package for REDUCE System
\cite{reduce}. The result of the calculation is the following. Making use of
the gauge condition (\ref{gauge}), one can reduce the functional structure of
each diagram to $p^2 h(-p) \phi^2(p),$
the coefficients being polynomials\footnote{The subscripts of the polynomials
refer to the corresponding diagrams of Figs.~\ref{fig1}-\ref{fig3}.}
on the gauge parameters $\alpha \equiv (\beta - 1)^{-1}$ and $\xi$
\begin{eqnarray}&&\label{divalt}
I(p) = \left\{- \frac{1}{2} m^2 p^2 h(-p) \phi^2(p)\right\}
\left(\frac{\mu^2}{p^2}\right)^{\varepsilon/2}
\frac{P(\alpha,\xi)}{8\pi^2\varepsilon},
\end{eqnarray}
where
\begin{eqnarray}&&
P_{1a}(\alpha,\xi) =
 - \frac{1}{4} \alpha^3 \xi^2 + \frac{3}{4} \alpha^3 \xi
+ \frac{3}{2} \alpha^2 \xi
+ \frac{3}{4} \alpha \xi - \frac{3}{4} \alpha^{-1} \xi^2,
\nonumber
\end{eqnarray}
\begin{eqnarray}&&
P_{1b}(\alpha,\xi) =
 - \frac{5}{32} \alpha^4 \xi^2 + \frac{15}{32} \alpha^4 \xi
- \frac{3}{32} \alpha^3 \xi^2 + \frac{29}{96} \alpha^3 \xi
\nonumber\\&&
+ \frac{35}{96} \alpha^2 \xi^2 - \frac{103}{96} \alpha^2 \xi
+ \frac{15}{32} \alpha \xi^2 - \frac{75}{32} \alpha \xi - \frac{7}{16} \xi,
\nonumber
\end{eqnarray}
\begin{eqnarray}&&
P_{1c}(\alpha,\xi) =
 - \frac{55}{96} \alpha^4 \xi^2 + \frac{55}{32} \alpha^4 \xi
- \frac{3}{32} \alpha^3 \xi^2 + \frac{253}{96} \alpha^3 \xi
\nonumber\\&&
+ \frac{35}{96} \alpha^2 \xi^2 - \frac{19}{96} \alpha^2 \xi
+ \frac{15}{32} \alpha \xi^2 - \frac{75}{32} \alpha \xi
- \frac{5}{4} \xi^2 - \frac{17}{16} \xi,
\nonumber
\end{eqnarray}
\begin{eqnarray}&&
P_{1d+2c+2d}(\alpha,\xi) =
 - \frac{5}{16} \alpha^4 \xi^2 + \frac{15}{16} \alpha^4 \xi
- \frac{3}{16} \alpha^3 \xi^2
\nonumber\\&&
+ \frac{125}{48} \alpha^3 \xi
+ \frac{35}{48} \alpha^2 \xi^2 + \frac{127}{48} \alpha^2 \xi
+ \frac{15}{16} \alpha \xi^2 + \frac{31}{48} \alpha \xi,
\nonumber
\end{eqnarray}
\begin{eqnarray}&&
P_{2a}(\alpha,\xi) =
 - \frac{5}{12} \alpha^4 \xi^2 + \frac{55}{32} \alpha^4 \xi
- \frac{1}{4} \alpha^3 \xi^2 + \frac{361}{96} \alpha^3 \xi
\nonumber\\&&
+ \frac{161}{96} \alpha^2 \xi
- \frac{221}{96} \alpha \xi
- \frac{3}{4} \alpha^{-1} \xi^2 - \frac{7}{12} \alpha^{-1} \xi
- \frac{5}{4} \xi^2 - \frac{137}{48} \xi,
\nonumber
\end{eqnarray}
\begin{eqnarray}&&
P_{2b}(\alpha,\xi) =
\frac{15}{32} \alpha^4 \xi + \frac{41}{96} \alpha^3 \xi
- \frac{131}{96} \alpha^2 \xi
\nonumber\\&&
- \frac{361}{96} \alpha \xi
- \frac{7}{12} \alpha^{-1} \xi - \frac{133}{48} \xi,
\nonumber
\end{eqnarray}
\begin{eqnarray}&&
P_{3a}(\alpha,\xi) =
\frac{1}{8} \alpha^3 \xi + \frac{11}{24} \alpha^2 \xi
+ \frac{1}{12} \alpha \xi,
\nonumber
\end{eqnarray}
\begin{eqnarray}&&
P_{3b}(\alpha,\xi) =
 - \frac{1}{8} \alpha^3 \xi - \frac{5}{24} \alpha^2 \xi
- \frac{5}{8} \alpha \xi - \frac{13}{24} \xi.
\nonumber
\end{eqnarray}

Finally, summing up individual contributions, taking into account
$$\Box h = 2\alpha R = 2\alpha T, ~~T = T^{\mu\nu}\eta_{\mu\nu},$$
and going over to coordinate space we have for the total gauge
dependence of the effective mass term of the scalar field action
\begin{eqnarray}&&\label{finalf}
\frac{d\Gamma_{\phi}}{d\xi} =
 \left\{- \int d^4 x \frac{1}{2} m^2\phi^2(x) \ln \Box ~T(x) \right\}
\frac{P^{tot}(\alpha,\xi)}{16 \pi^2},
\end{eqnarray}
where
\begin{eqnarray}&&
P^{tot}(\alpha,\xi) = \left( \frac{35}{24} \alpha^5 \xi
- \frac{85}{16} \alpha^5 + \frac{7}{8} \alpha^4 \xi
- \frac{503}{48} \alpha^4 - \frac{35}{24} \alpha^3 \xi
\right.
\nonumber\\&&
\left.
- \frac{55}{16} \alpha^3 - \frac{15}{8} \alpha^2 \xi
+ \frac{475}{48} \alpha^2 + \frac{5}{2}\alpha \xi
+ \frac{23}{3}\alpha + \frac{3}{2} \xi + \frac{7}{6} \right)
\nonumber
\end{eqnarray}
Although we have only determined gauge dependence of the
effective mass term, this is enough to conclude that the effective
equations of motion of the apparatus depend on the choice of the gauge.
One might think that these equations could still turn out to be
gauge-independent provided that the effective kinetic term had an
appropriate gauge-dependent part, e.g., the one described by equation
analogous to Eq.~(\ref{finalf}) with the same polynomial
$P^{tot}(\alpha,\xi).$ Remember, however, that besides the terms of the
order $\hbar^{1},$ whose $\xi$-dependence is described by
Eq.~(\ref{finalf}), the effective apparatus action contains also
terms of the order $\hbar^{0},$ corresponding to the classical (tree)
approximation. The latter are of course gauge-independent, so there
can be no cancellation of the gauge-dependent factors in the effective
equations of motion. Obviously, no manipulation with these equations
can change this conclusion, since Eq.~(\ref{finalf}) alone is already
sufficient to determine the gauge dependence of the effective
gravitational field. Indeed, as explained in the Introduction,
this field is to be determined by comparison of the effective and
classical equations of motion of the measuring device
(or by comparison of the corresponding action functionals).
The mass term in the action (\ref{actionm}) for the scalar particle
in a weak gravitational field is
\begin{eqnarray}&&
- \int d^4 x \frac{1}{2} m^2\phi^2(x) \frac{h_{\mu\nu}(x) \eta^{\mu\nu}}{2}.
\nonumber
\end{eqnarray}
Therefore, in view of arbitrariness of the source $T^{\mu\nu},$ we
conclude that the gauge dependence of the effective gravitational
field measured by means of the scalar particle is described by the following
equation
\begin{eqnarray}&&\label{final}
\frac{d h_{\mu\nu}^{eff}(x)}{d\xi} =
\ln \Box ~T_{\mu\nu}(x) \frac{P^{tot}(\alpha,\xi)}{8 \pi^2}.
\end{eqnarray}
Thus, unlike the case of the point-like measuring apparatus,
considered in \cite{dalvit}, the value of the effective gravitational field
measured by the scalar field turns out to be gauge-dependent.
However, the non-vanishing of the right hand side of Eq.~(\ref{final})
should not discourage, since as we have mentioned in the Introduction this
result does not take into account gauge dependence of the radiation
corrections to the classical form of the scalar field action (i.e., the
gravitational form factors of the scalar particle), which in the case of
the gravitational interaction do not disappear in the limit $p^2/m^2 \to 0.$

\section{Conclusion}\label{conclud}

The results of the calculation presented in this paper show that
from the field-theoretical point of view, explicit inclusion of the
measuring apparatus into mathematical framework of the quantum
theory does not solve the problem of gauge dependence of observables,
unlike the case when the effective field is measured by point particles,
discussed in \cite{dalvit}.

It must be emphasized, however, that we have obtained this result
under assumption that the action for the measuring device can be
taken in the ordinary classical form. Unlike all other fundamental
interactions, in the case of gravity this choice of the action
cannot be justified by the standard limiting arguments \cite{donoghue}.
In other words, consistent field-theoretical approach requires
taking into account quantum corrections to the classical form
of the device action, i.e., evaluation of the corresponding form factors.
Such evaluation is by itself very complicated task, since one has to deal
with Feynman graphs whose external lines all are off the mass-shell.
Its results would be decisive for answering the question of whether
approach suggested in \cite{dalvit} does solve the gauge dependence problem.

Leaving this question open we would like to note only that its eventual
resolution would serve as a valuable guide in investigation of the
fundamental role played by the measurement process in quantum theory,
of which description presented above is just a phenomenology.

\vspace{1cm}
{\Large \bf Appendix}
\vspace{1cm}

As we saw in Sec.~\ref{slavid}, the effective Slavnov identities
for the combined system "gravitational field plus measuring device"
contain the second derivatives of the generating functionals
of Green functions with respect to the sources. If the apparatus
contribution to the effective gravitational field is infinitesimal,
this fact is inessential since then evaluation of the gauge
dependence of the generating functional of simply connected Green functions
is needed only [see Eq.~(\ref{main})]. However, in the general
case of finite device contribution, one has to deal with the
effective Slavnov identity for the generating functional of
one-particle-irreducible Green functions, i.e., effective action.
In this case, one can avoid the complications caused by the presence
of the second derivatives, which arise when one performs the Legendre
transformation in Eq.~(\ref{slav2}). This can be done as follows.

As seen from the derivation of identities (\ref{slav1}), (\ref{slav2}),
given in Sec.~\ref{slavid}, appearance of the second derivatives
in these identities is traced to non-invariance of the apparatus
action with respect to the "quantum" BRST transformation (\ref{brsta})
of integration variables in the generating functional (\ref{genernew}).
The first term in the square brackets in Eq.~(\ref{slav}),
representing this non-invariance, is non-linear on the quantum fields.
On the other hand, the standard source for the BRST transformation of the
field $\phi$ (the Zinn-Justin source, \cite{zinnjustin}) is linear
on the quantum fields, since the generator $\tilde{D}(\phi)$ is
a functional of the classical field $\phi$ only. In this sense, introduction
of the source $\tilde{K}$ into the quantum action is superfluous.
It is natural, therefore, to introduce a source for the BRST variation
of the whole device action, instead of the source $\tilde{K}.$
Namely, let us consider the generating functional (\ref{genernew}),
where $\Sigma$ stands for
\begin{eqnarray}&&
\Sigma = S_{fp} + K^{\mu\nu}D_{\mu\nu}^{\alpha}C_{\alpha}
+ \frac{1}{2}L^{\gamma} f^{\alpha\beta}_{~~~\gamma}C_{\alpha}C_{\beta}
+ J \frac{\delta S_{\phi}}{\delta\phi_{i}}
\tilde{D}_{i}^{\alpha} C_{\alpha},
\nonumber
\end{eqnarray}
$J$ being the new constant anticommuting parameter.
Here we consider the general case when the measuring device is
described by an arbitrary number of fields $\phi_{i},$
denoting them collectively as $\phi.$
Obviously, the new source term is invariant under the "quantum"
BRST transformation (\ref{brsta}).

To derive the effective Slavnov identities, we do as in
the beginning of Sec.~\ref{slavid} and obtain the following identity

\begin{eqnarray}&&\label{slavalt}
{\displaystyle\int}dh dC d\bar{C}
\left[i \frac{\delta S_{\phi}}{\delta h_{\mu\nu}} D^{\alpha}_{\mu\nu} C_{\alpha}
+ i Y \bar{C}^{\alpha} F_{\alpha}^{,\mu\nu} D^{\beta}_{\mu\nu} C_{\beta}
+ ~i \frac{Y}{\xi} F_{\alpha}^{2}
+ T^{\mu\nu} \frac{\delta}{\delta K^{\mu\nu}}
\right.
\nonumber\\&&
\left.
- \bar{\beta}^{\alpha}\frac{\delta}{\delta L^{\alpha}}
- i \beta_{\alpha}\frac{F^{\alpha}}{\xi}
\right]
\exp\{i (\Sigma
+ Y F_{\alpha}\bar{C}^{\alpha}
+ \bar{\beta}^{\alpha}C_{\alpha} + \bar{C}^{\alpha}\beta_{\alpha}
+ T^{\mu\nu}h_{\mu\nu})\} = 0.
\end{eqnarray}

The first term in the square brackets is now replaced simply by the
minus derivative with respect to the source $J.$
Applying the ghost equation of motion to the second term, we rewrite
Eq.~(\ref{slavalt}) as
\begin{eqnarray}&&
\left( T^{\mu\nu}\frac{\delta}{\delta K^{\mu\nu}}
- \bar{\beta}^{\alpha}\frac{\delta}{\delta L^{\alpha}}
- \frac{1}{\xi} \beta_{\alpha}F^{\alpha,\mu\nu}\frac{\delta}{\delta T^{\mu\nu}}
- \frac{\partial}{\partial J}
- Y \beta_{\gamma}\frac{\delta}{\delta\beta_{\gamma}}
- 2 Y\xi\frac{\partial}{\partial\xi}
\right) Z  = 0,
\nonumber
\end{eqnarray}
which in terms of the generating functional of connected Green functions
takes the form
\begin{eqnarray}\label{slav2alt}&&
T^{\mu\nu}\frac{\delta \tilde{W}}{\delta K^{\mu\nu}}
- \bar{\beta}^{\alpha}\frac{\delta \tilde{W}}{\delta L^{\alpha}}
- \frac{1}{\xi} \beta_{\alpha}F^{\alpha,\mu\nu}\frac{\delta \tilde{W}}{\delta T^{\mu\nu}}
- \frac{\partial \tilde{W}}{\partial J}
- Y \beta_{\alpha}\frac{\delta \tilde{W}}{\delta\beta_{\alpha}}
- 2 Y\xi\frac{\partial \tilde{W}}{\partial\xi} = 0.
\end{eqnarray}
Now, the Legendre transformation is easily performed
in Eq.~(\ref{slav2alt}):
with the help of equations
\begin{eqnarray}&&\label{meanhinv}
T^{\mu\nu} =  - \frac{\delta \tilde{\Gamma}}{\delta h_{\mu\nu}},
~~\bar{\beta}^{\alpha} = \frac{\delta \tilde{\Gamma}}{\delta C_{\alpha}},
~~\beta_{\alpha} = - \frac{\delta \tilde{\Gamma}}{\delta\bar{C}^{\alpha}},
\nonumber
\end{eqnarray}
which are the inverse of Eqs.~(\ref{meanh}) -- (\ref{meanbc}),
and the relations
\begin{eqnarray}
\frac{\delta \tilde{W}}{\delta K^{\mu\nu}} = \frac{\delta \tilde{\Gamma}}{\delta K^{\mu\nu}},
~~\frac{\delta \tilde{W}}{\delta \xi} = \frac{\delta \tilde{\Gamma}}{\delta \xi}, {\rm ~~etc.}
\nonumber
\end{eqnarray}
we bring Eq.~(\ref{slav2alt}) to
\begin{eqnarray}\label{slav3alt}&&
\frac{\delta \tilde{\Gamma}}{\delta h_{\mu\nu}}\frac{\delta \tilde{\Gamma}}{\delta K^{\mu\nu}}
+ \frac{\delta \tilde{\Gamma}}{\delta C_{\alpha}}\frac{\delta \tilde{\Gamma}}{\delta L^{\alpha}}
- \frac{F^{\alpha}}{\xi} \frac{\delta \tilde{\Gamma}}{\delta\bar{C}^{\alpha}}
+ \frac{\partial \tilde{\Gamma}}{\partial J}
+ Y \frac{\delta \tilde{\Gamma}}{\delta\bar{C}^{\alpha}}\bar{C}^{\alpha}
+ 2 Y\xi\frac{\partial \tilde{\Gamma}}{\partial\xi} = 0.
\end{eqnarray}
This is desired effective Slavnov identity for the generating functional
of one-particle-irreducible Green functions. Also, it is easy to verify
that upon extraction of the $\phi$-dependent terms in Eq.~(\ref{slav2alt}),
the (one-loop) functional $\partial\tilde{W}/\partial J$ generates just the
set of diagrams in Figs.~\ref{fig2}, \ref{fig3}.

\end{document}